\documentclass[letterpaper,twocolumn,10pt]{article}
  \usepackage{usenix2019_v3}

  \usepackage{amsmath}
  \usepackage{graphicx}
  \usepackage{xcolor}
  \usepackage{booktabs}
  \usepackage{array}
  \usepackage{url}
  \usepackage{setspace}
 \usepackage{subcaption}
  \usepackage{dblfloatfix}   

\newcommand{\fixme}[1]{{\color{red}\textbf{\fbox{FIXME} #1}}}

\renewcommand{\em}{\it}

\newcommand{\ignore}[1]{}

\def\cfigure[#1,#2,#3]{
\begin{figure}
\vspace*{0mm}
\begin{center}

\includegraphics[width=3in]{#1} 
 
\vspace*{-3mm}\caption[]{#2
} \label{#3}
 
\vspace*{-5mm}
\end{center}
\end{figure}}

\def\cfigurefour[#1,#2,#3]{
\begin{figure}
\vspace*{0mm}
\begin{center}

\includegraphics[width=4in]{#1} 
 
\vspace*{-3mm}\caption[]{#2
} \label{#3}
 
\vspace*{-5mm}
\end{center}
\end{figure}}

\def\cfiguretemp[#1,#2,#3]{
\begin{figure}
\vspace*{0mm}
\begin{center}

\includegraphics[width=3.5in]{#1} 
 
\vspace*{-3mm}\caption[]{#2
} \label{#3}
 
\vspace*{-5mm}
\end{center}
\vspace*{-2mm}
\end{figure}}

\def\wfigure[#1,#2,#3]{
\begin{figure*}
\vspace*{0mm}
\begin{center}
 \includegraphics[width=\textwidth]{#1} 
 \vspace*{-3mm}\caption[]{#2
} \label{#3}
 
\end{center}
\end{figure*}}

\def\threefigure[#1,#2,#3,#4,#5]{
\begin{figure*}
\vspace*{0mm}
\begin{center}

\begin{tabular}{ccc}
\includegraphics[width=2in]{#1} & \includegraphics[width=2in]{#2} &  \includegraphics[width=2in]{#3} \\
(a) & (b) & (c) \\
\end{tabular}

\vspace*{-3mm}\caption[]{#4
} \label{#5}

\vspace*{-5mm}
\end{center}
\vspace*{-2mm}
\end{figure*}}

\def\dcfigure[#1,#2,#3,#4,#5,#6]{
{
\begin{figure*}
\begin{center}
\begin{minipage}[c]{\columnwidth}{
\includegraphics[width=\columnwidth]{#1} 
\vspace*{0mm}\caption[]{#2} \label{#3} \
}\end{minipage}\hspace*{\columnsep}\
\begin{minipage}[c]{\columnwidth}{
\includegraphics[width=\columnwidth]{#4} 
\vspace*{0mm}\caption[]{#5}\label{#6} \
}\end{minipage}
\end{center}
\end{figure*}
}
}

\def\tableByTable[#1,#2,#3,#4,#5,#6]{
{
\begin{table*}
\begin{center}
\begin{minipage}[c]{3in}{
\centering
{#1}
\vspace*{0mm}\tabcaption[]{#2}\label{#3} \
}\end{minipage}\hspace*{\columnsep}\
\begin{minipage}[c]{3in}{
\centering
{#4}
\vspace*{0mm}\tabcaption[]{#5}\label{#6} \
}\end{minipage}
\end{center}
\end{table*}
}
}

\def\figureByTable[#1,#2,#3,#4,#5,#6]{
{
\begin{figure*}
\begin{center}
\begin{minipage}[c]{3in}{
\centering
\includegraphics[width=\textwidth]{#1}
\vspace*{0mm}\figcaption[]{#2} \label{#3} \
}\end{minipage}\hspace*{\columnsep}\
\begin{minipage}[c]{3.3in}{
\centering
{#4}
\vspace*{0mm}\tabcaption[]{#5}\label{#6} \
}\end{minipage}
\end{center}
\end{figure*}
}
}

\def\tableByFigure[#1,#2,#3,#4,#5,#6]{
{
\begin{figure*}
\begin{center}
\begin{minipage}[c]{4.3in}{
\centering
{#1}
\vspace*{0mm}\tabcaption[]{#2} \label{#3} \
}\end{minipage}\hspace*{\columnsep}\
\begin{minipage}[c]{2.2in}{
\centering
\includegraphics[width=\textwidth]{#4}
\vspace*{-0.35in}\caption[]{#5}\label{#6} \
}\end{minipage}
\end{center}
\end{figure*}
}
}

\def\doublecfigure[#1,#2,#3,#4]{
{
\begin{figure}
\begin{center}
\begin{minipage}[c]{1.5in}{
\begin{center}
\includegraphics[width=1.5in]{#1}
\end{center}
}\end{minipage}\hspace*{1em}\
\begin{minipage}[c]{1.5in}{
\begin{center}
\includegraphics[width=1.5in]{#2}
\end{center}
}\end{minipage}
\vspace*{0mm}\caption[]{#3} \label{#4} \
\end{center}
\end{figure}
}
}

\def\qcfigure[#1,#2,#3,#4,#5,#6]{
{
\begin{figure*}
\vspace*{0.2in}\
\begin{center}
\begin{minipage}[c]{3in}{
\includegraphics[width=3in]{#1} 
\vspace*{-3mm}
}
\end{minipage}\hspace*{0.5in}\
\begin{minipage}[c]{3in}{
\includegraphics[width=3in]{#2} 
\vspace*{-3mm}
}\end{minipage}

\begin{minipage}[c]{3in}{
\includegraphics[width=3in]{#3} 
\vspace*{-3mm}
}
\end{minipage}\hspace*{0.5in}\
\begin{minipage}[c]{3in}{
\includegraphics[width=3in]{#4} 
\vspace*{-3mm}
}\end{minipage}
\end{center}
\caption[]{#5}\label{#6}
\end{figure*}
}
}

\def\twfigure[#1,#2,#3,#4,#5]{
{
\begin{figure*}
\vspace*{0.2in}\
\begin{center}
\begin{minipage}[c]{6.5in}{
\includegraphics[width=6.5in]{#1} 
\vspace*{-3mm}
}
\end{minipage}

\begin{minipage}[c]{6.5in}{
\includegraphics[width=6.5in]{#2} 
\vspace*{-3mm}
}\end{minipage}

\begin{minipage}[c]{6.5in}{
\includegraphics[width=6.5in]{#3} 
\vspace*{-3mm}
}
\end{minipage}
\end{center}
\caption[]{#4}\label{#5}
\end{figure*}
}
}

\def\dwfigure[#1,#2,#3,#4]{
{
\begin{figure*}
\vspace*{0.2in}\
\begin{center}
\begin{minipage}[c]{6.5in}{
\includegraphics[width=6.5in]{#1} 
\vspace*{-3mm}
}
\end{minipage}

\begin{minipage}[c]{6.5in}{
\includegraphics[width=6.5in]{#2} 
\vspace*{-3mm}
}\end{minipage}

\end{center}
\caption[]{#3}\label{#4}
\end{figure*}
}
}

\def\dssfigure[#1,#2,#3,#4,#5,#6]{
{
\begin{figure*}
\vspace*{0.2in}\
\begin{center}
\begin{minipage}[c]{4in}{
\includegraphics[width=4in]{#1}
\vspace*{-3mm}\caption[]{#2} \label{#3} \
}\end{minipage}\hspace*{0.5in}\
\begin{minipage}[c]{2in}{
\includegraphics[width=2in]{#4}
\vspace*{-3mm}\caption[]{#5}\label{#6} \
}\end{minipage}
\end{center}
\vspace*{-0.4in}\
\end{figure*}
}
}

\def\dsfigure[#1,#2,#3,#4,#5,#6]{
{
\begin{figure*}
\vspace*{0.2in}\
\begin{center}
\begin{minipage}[c]{3in}{
\includegraphics[width=3in]{#1}
\vspace*{-3mm}\caption[]{#2} \label{#3} \
}\end{minipage}\hspace*{0.5in}\
\begin{minipage}[c]{3in}{
\hspace*{0.5in}\
\includegraphics[height=3in]{#4}
\vspace*{-3mm}\caption[]{#5}\label{#6} \
}\end{minipage}
\end{center}
\vspace*{-0.4in}\
\end{figure*}
}
}

\def\dsyfigure[#1,#2,#3,#4,#5,#6]{
{
\begin{figure*}
\vspace*{0.2in}\
\begin{center}
\begin{minipage}[c]{2.5in}{
\includegraphics[height=2.5in]{#1}
\vspace*{-3mm}\caption[]{#2} \label{#3} \
}\end{minipage}\hspace*{0.5in}\
\begin{minipage}[c]{2.5in}{
\includegraphics[height=2.5in]{#4}
\vspace*{-3mm}\caption[]{#5}\label{#6} \
}\end{minipage}
\end{center}
\vspace*{-0.4in}\
\end{figure*}
}
}

\def\dyfigure[#1,#2,#3,#4,#5,#6]{
{
\begin{figure*}
\vspace*{0.2in}\
\begin{center}
\begin{minipage}[c]{3in}{
\includegraphics[height=3in]{#1} 
\vspace*{-3mm}\caption[]{#2} \label{#3} \
}\end{minipage}\hspace*{0.5in}\
\begin{minipage}[c]{3in}{
\includegraphics[height=3in]{#4} 
\vspace*{-3mm}\caption[]{#5}\label{#6} \
}\end{minipage}
\end{center}
\vspace*{-0.4in}\
\end{figure*}
}
}

\def\dyoldfigure[#1,#2,#3,#4,#5,#6]{
{
\begin{figure*}
\vspace*{0.2in}\
\begin{center}
\begin{minipage}[c]{3in}{
\epsfysize=2.0in\
\hspace{0.5in}\
\epsfbox{#1}
\vspace*{-3mm}\caption[]{#2} \label{#3} \
}\end{minipage}\hspace*{0.25in}\
\begin{minipage}[c]{3in}{
\epsfysize=2.0in\
\hspace{0.5in}\
\epsfbox{#4}
\vspace*{-3mm}\caption[]{#5}\label{#6} \
}\end{minipage}
\end{center}
\vspace*{-0.4in}\
\end{figure*}
}
}

\def\cfiguredouble[#1,#2,#3,#4]{
\begin{figure}
\vspace*{0.2in}\
\begin{center}
\begin{minipage}[c]{1.5in}{
\epsfxsize=1.5in\
\epsfbox{#1}
}\end{minipage}\hspace*{0.1in}\
\begin{minipage}[c]{1.5in}{
\epsfxsize=1.5in\
\vspace{0.1in}\epsfbox{#2}
}\end{minipage}\vspace*{-0.10in} \caption[]{#3}\label{#4}
\end{center}
\vspace*{-0.4in}\
\end{figure}
}

\def\wpfigure[#1,#2,#3,#4]{
\begin{figure*}
\vspace*{4mm}
\begin{center}

\includegraphics[width=#4]{#1} 

\vspace*{-3mm}\caption[]{#2
} \label{#3}

\vspace*{-5mm}
\end{center}
\end{figure*}}

\def\wprfigure[#1,#2,#3,#4,#5]{
\begin{figure*}
\vspace*{4mm}
\begin{center}

\includegraphics[width=#4, angle=#5]{#1} 

\vspace*{-3mm}\caption[]{#2
} \label{#3}

\vspace*{-5mm}
\end{center}
\end{figure*}}

\def\DoubleFigureWSlide[#1,#2,#3,#4,#5,#6,#7,#8,#9]{
\begin{figure*}
\vspace*{#9}
\begin{center}
\begin{minipage}{#4}
\includegraphics[width=#4]{#1}
\vspace*{-3mm}\caption{#2
}\label{#3}
\end{minipage}
\hspace{2em}
\begin{minipage}{#8}
\includegraphics[width=#8]{#5}
\vspace*{-3mm}\caption{#6
}\label{#7}
\end{minipage}
\vspace*{-5mm}
\end{center}
\end{figure*}
}

\def\DoubleFigureW[#1,#2,#3,#4,#5,#6,#7,#8]{
\begin{figure*}
\vspace*{0in}
\begin{center}
\begin{minipage}{#4}
\includegraphics[width=#4]{#1}
\vspace*{-3mm}\caption{#2
}\label{#3}
\end{minipage}
\hspace{2em}
\begin{minipage}{#8}
\includegraphics[width=#8]{#5}
\vspace*{-3mm}\caption{#6
}\label{#7}
\end{minipage}
\vspace*{-5mm}
\end{center}
\end{figure*}
}

\def\DoubleFigureWHack[#1,#2,#3,#4,#5,#6,#7,#8]{
\begin{figure*}
\vspace*{0in}
\begin{center}
\begin{minipage}{3in}
\includegraphics[width=#4]{#1}
\vspace*{-3mm}\caption{#2
}\label{#3}
\end{minipage}
\hspace{2em}
\begin{minipage}{3in}
\includegraphics[width=#8]{#5}
\vspace*{-3mm}\caption{#6
}\label{#7}
\end{minipage}
\vspace*{-5mm}
\end{center}
\end{figure*}
}

\def\ddcfigure[#1,#2,#3,#4]{
\begin{figure*}
\vspace*{0.2in}\
\begin{center}
\begin{minipage}[c]{\columnwidth}{
\includegraphics[width=\columnwidth]{#1} 
}\end{minipage}\hspace{0.5in}\
\begin{minipage}[c]{\columnwidth}{
\includegraphics[width=\columnwidth]{#2} 
}\end{minipage} \caption[]{#3}\label{#4}
\end{center}
\end{figure*}
}

\def\ddcfigureSlide[#1,#2,#3,#4,#5]{
\begin{figure*}
\vspace*{#5}\
\begin{center}
\begin{minipage}[c]{3in}{
\includegraphics[height=3in]{#1} 
}\end{minipage}\hspace{0.5in}\
\begin{minipage}[c]{3in}{
\includegraphics[height=3in]{#2} 
}\end{minipage}\vspace*{-0.10in} \caption[]{#3}\label{#4}
\end{center}
\vspace*{-0.4in}\
\end{figure*}
}

\def\cxfigure[#1,#2,#3]{
\begin{figure}
\vspace*{4mm}
\begin{center}
 
\epsfxsize=2.5in\
\epsfbox{#1}\
 
\vspace*{-0.10in}\caption[]{#2
} \label{#3}
 
\vspace*{-5mm}
\end{center}
\vspace*{-2mm}
\end{figure}}

\newcommand{\beforecaption}{\vspace{-.15cm}\begin{spacing}{0.85}}
\newcommand{\aftercaption}{\vspace{-.45cm}\end{spacing}}
\newcommand{\mycaption}[3]{\beforecaption\caption{\label{#1}{\bf #2} \em\small #3}\aftercaption}

\newcommand{\eg}{\textit{e.g.}}
\newcommand{\ie}{\textit{i.e.}}

\newcommand{\GB}{\,GB}

\newcommand{\ms}{\mbox{$ms$}}

\newcommand{\boldunderpara}[1]{\noindent{\underline{\textbf{#1}}}}

\newcommand{\sys}{Nitsum}

\usepackage{tikz}

\begin{document}
  \raggedbottom
  \date{}

  \title{\Large \bf \sys: Serving Tiered LLM Requests with Adaptive Tensor Parallelism}

\author{
{\rm Vikranth Srivatsa}\\
University of California, San Diego
\and
{\rm Zijian He}\\
University of California, San Diego
\and
{\rm Pu Guo}\\
University of California, San Diego
\and
{\rm Dongming Li}\\
University of California, San Diego
\and
{\rm Yiying Zhang}\\
University of California, San Diego and GenseeAI Inc.
}
  \maketitle

  \begin{abstract}
LLM serving is increasingly multi-tenant: the same deployment must handle latency-critical interactive requests and more relaxed background workloads under a fixed GPU budget. This creates a tiered-SLO setting where maximizing overall goodput (requests that satisfy both TTFT and TPOT targets) is challenging because workload mix, request lengths, and load intensity vary over time. Existing systems mainly optimize request-level controls (e.g., queuing and batching) while keeping execution configuration largely static, which limits adaptation under multi-tier contention.

We present \textbf{\sys}, a distributed LLM serving system that treats tensor parallelism (TP) as a first-class \emph{runtime} control surface rather than a static deployment choice. \sys{} jointly optimizes TP level, prefill/decode GPU split, and request scheduling. To make frequent TP adaptation practical, \sys introduces TP-aware weight reuse and fast KV migration.
Experiments on real traces and targeted microbenchmarks show that \sys{} improves SLO-compliant goodput over SoTA by up to 5.3 times.
\end{abstract}

  \section{Introduction}
\label{sec:intro}

LLM serving is increasingly tiered. A single model deployment~\cite{claude-opus-4.6} now serves interactive chat~\cite{claude-chat-interface}, coding agents~\cite{claude-code}, computer-use agents~\cite{claude-cowork}, API calls embedded in products~\cite{claude-api}, and long-running background or scheduled jobs~\cite{does-claude-have-background-pricing} on the same infrastructure. These requests have very different latency expectations: some require fast first-token and steady token rates for user interaction, while others tolerate slower responses in exchange for lower cost~\cite{lowlatency,convo-latency,openai2023batch,databricks-batch}. In practice, this creates \emph{tiers} of service objectives.

Providing tiered-SLO serving would be straightforward if GPU resources were abundant: one could simply provision a dedicated cluster with max capacity for each SLO tier. In practice, however, model providers operate under fixed GPU budgets, while tier mix, request lengths, and load intensity vary substantially over time~\cite{azure-trace,alibaba-trace}. Strictly separating clusters would therefore waste substantial capacity, while pooling all requests together creates interference among heterogeneous TTFT and TPOT objectives. 
This leads to the central question of this paper:

\begin{quote}
\itshape
How should an LLM serving system operate under a fixed GPU budget and maximize the number of requests per second that meet both their TTFT and TPOT SLOs (\ie, goodput) for multiple SLO tiers?
\end{quote}

Existing SLO-aware LLM serving systems provide part of the answer. Systems such as QLM~\cite{onequeue}, Llumnix~\cite{llumnix}, Chiron~\cite{chiron}, and SCOOT~\cite{scoot} support heterogeneous SLOs through request-level control mechanisms, including queuing, batching, migration, and autoscaling. These techniques are effective at prioritizing requests and reducing queueing delay.
However, they primarily control \textbf{\textit{when}} requests run, not \textbf{\textit{how}} they are executed. The execution configuration of each request, and thus its TTFT and TPOT behavior, remains largely fixed. As a result, these approaches cannot directly change the per-request service time that ultimately determines SLO attainment. 



We make a key observation: the model execution configuration itself can be used to improve SLO attainment. In particular, \textbf{Tensor Parallelism (TP)}, typically treated as a static knob for fitting models across GPUs, also has a direct impact on \emph{SLO-relevant serving performance}. Higher TP can reduce prefill latency, improving TTFT for requests with long prompts or tight first-token targets. More surprisingly, we find that higher TP can also improve \emph{decode throughput} in low-batch-size regimes (but not high-batch-size ones) because of an intra-GPU cache and inter-GPU communication bandwidth tradeoff. This means TP can influence both TTFT and TPOT behavior, making it a practical control surface for LLM serving SLOs rather than merely a deployment choice.

\begin{figure}
\begin{center}
\centerline{\includegraphics[width=\columnwidth]{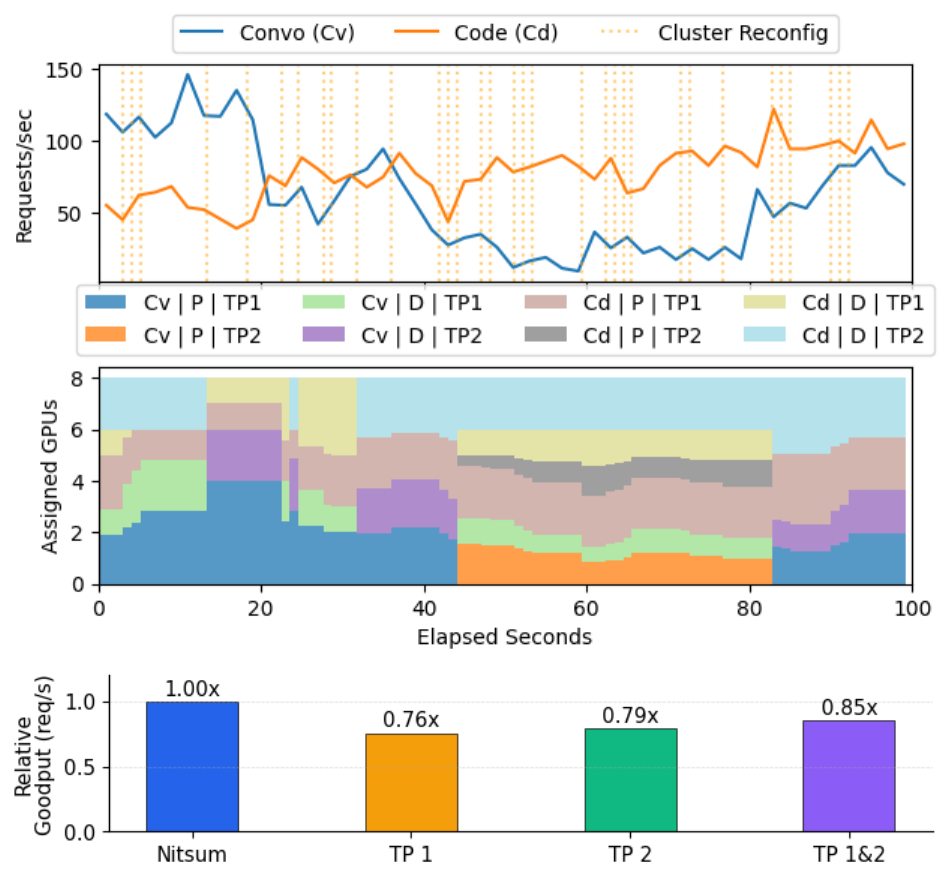}}
\mycaption{fig-intro-motiv}{Tiered SLO Workload and Cluster Dynamism.}
{
ServeGen~\cite{alibaba-trace} conversation and coding workloads running on 8 H100 GPUs and Llama 3.1-8B. Top row: request pattern and cluster reconfiguration. Middle: optimal cluster configuration of ``request group | Prefill/Decode stage | TP level''. Bottom: \sys{} and static TP configuration goodput (higher is better)
. 
}
\end{center}
\end{figure}


Our observation has an important system-level implication. Because TP affects prefill and decode differently, and because different SLO tiers impose different latency pressure, the execution configuration that maximizes SLO-compliant goodput is not fixed. As workload mix and service pressure change over time, the goodput-optimal configuration can shift substantially. Figure~\ref{fig-intro-motiv} illustrates this effect: under a realistic workload from Alibaba cloud~\cite{alibaba-trace}, both request demand and optimal cluster configuration vary continuously, and no single static TP setting achieves the best performance. 


Clearly, systems that can adapt their execution configuration at runtime can achieve significantly higher goodput.
However, doing so is challenging in practice. Naively changing TP requires weight reloading, kernel reinitialization, and KV-cache movement across GPUs, which can take seconds to tens of seconds. 
These overheads are large relative to the duration of typical workload bursts, as shown in Figure~\ref{fig-intro-motiv}, and they can negate any potential benefit. In fact, from our experiments, today's system ends up spending most of its time reconfiguring rather than serving requests, effectively driving goodput close to zero.
Thus, the central challenge is to make execution-level TP adaptation fast enough to track workload dynamics.


In this paper, we present \textbf{\textit{\sys}}, a distributed LLM serving system that addresses this problem. \sys{} maximizes overall system goodput under a fixed GPU budget by jointly reconfiguring TP level, prefill/decode GPU allocation, and request placement at runtime, while maintaining fairness across SLO tiers. Notably, our cluster adaptation and scheduling policy treats TP reconfiguration as effectively cost-free, which greatly simplifies the optimization. Realizing this abstraction in practice requires reducing today's TP switching overhead from seconds or even tens of seconds to near zero. 

\sys{} introduces two key enabling mechanisms to achieve near-zero TP switching overhead. First, it eliminates model reloads during TP changes by storing one full weight copy per GPU and selecting TP-specific shards at execution time, while keeping TP-specialized execution processes warm. Second, it bounds TP-transition latency with an aggregated, pipelined KV migration mechanism that makes stop-and-migrate reconfiguration fast enough for sub-second-level control. With switching overhead reduced to this level, \sys{} can treat TP as an ``almost-free'' runtime knob rather than a static deployment choice.

On top of these mechanisms, \sys{} performs goodput-aware cluster reconfiguration and request scheduling. It uses offline prefill/decode profiles together with recent arrival statistics and tier-specific TTFT/TPOT SLOs to estimate the SLO-compliant throughput of candidate configurations. It then assigns GPUs across tiers and stages using a weighted greedy policy that balances goodput efficiency with fairness across tiers. The resulting configuration is enforced by global and local schedulers that coordinate request feasibility, placement, and iteration-level batch formation.

We evaluate \sys{} against four baselines: SGLang~\cite{sglang}, Llumnix~\cite{llumnix}, Chiron~\cite{chiron}, and a separated per-SLO-tier cluster setup. Our results on two GPU platforms, two LLMs, and two real traces~\cite{azure-trace,alibaba-trace} show that \sys{} consistently improves goodput over all baselines, by up to 5.3$\times$. In regimes where prior baselines collapse to near-zero goodput, \sys{} continues to sustain substantial SLO-compliant throughput.

In summary, this paper makes the following contributions:
\begin{itemize}
    \item Detailed analysis of how TP affects TTFT and TPOT for different request/batch sizes.
    \item The proposal of using TP as an effective control surface for meeting tiered TTFT/TPOT SLOs.
    \item Two mechanisms for low-overhead TP switching.
    \item A full set of policy for dynamic cluster reconfiguration and request scheduling for multi-tier-SLO goodput.
\end{itemize}

We will make \sys{} publicly available soon.
  \begin{figure*}[htbp]
  \centering
  \newcommand{\imgsize}{0.28\textwidth} 
  \setlength{\tabcolsep}{2pt} 

  \begin{tabular}{ccc}
    
    \includegraphics[width=\imgsize]{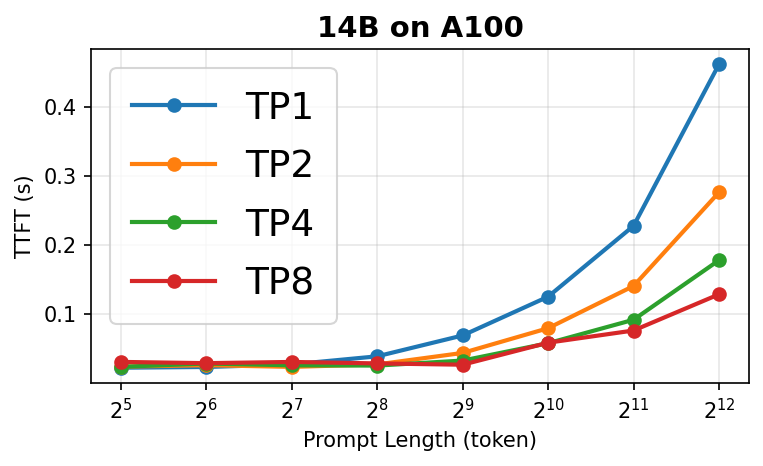} &
    \includegraphics[width=\imgsize]{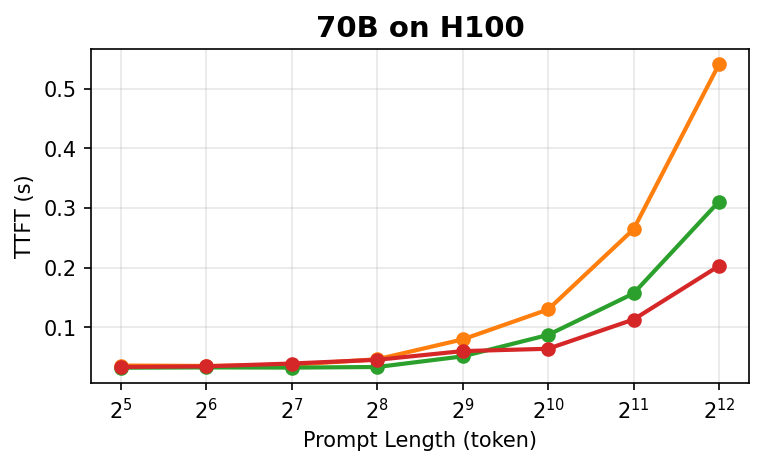} &
    \includegraphics[width=\imgsize]{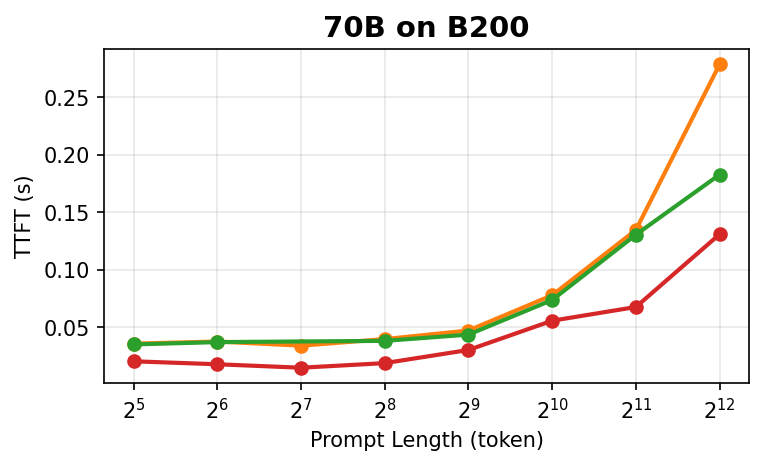} \\[-0.5ex]
    
    \includegraphics[width=\imgsize]{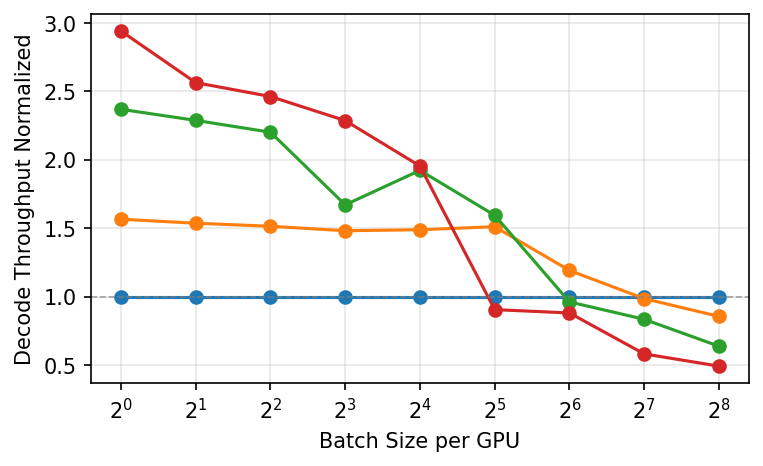} &
    \includegraphics[width=\imgsize]{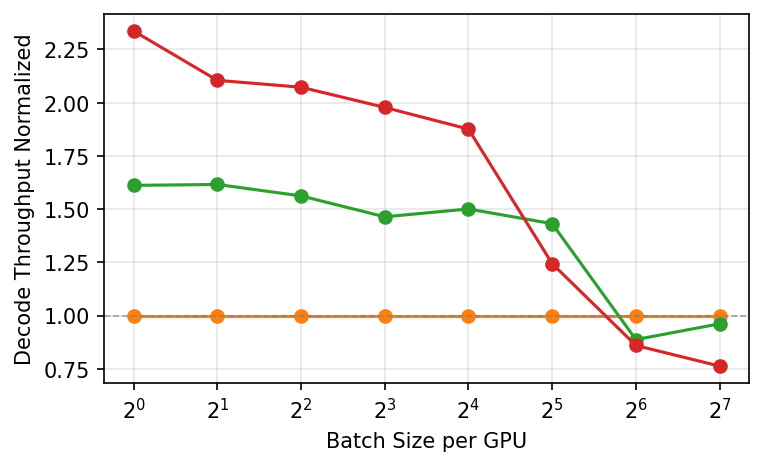} &
    \includegraphics[width=\imgsize]{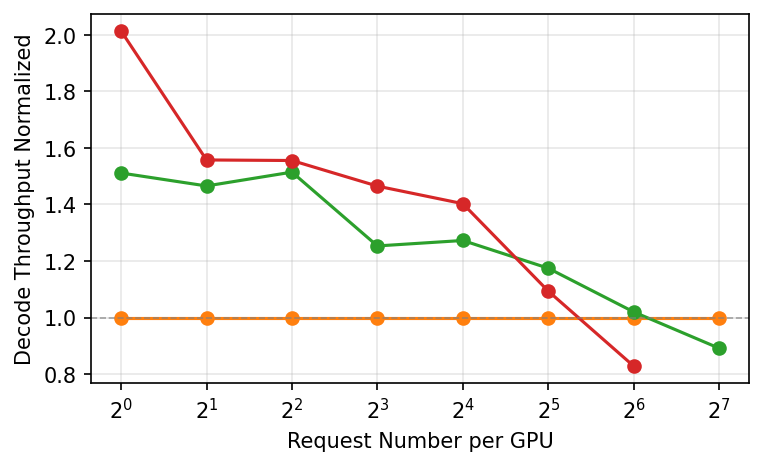} \\[-0.5ex]

    \includegraphics[width=\imgsize]{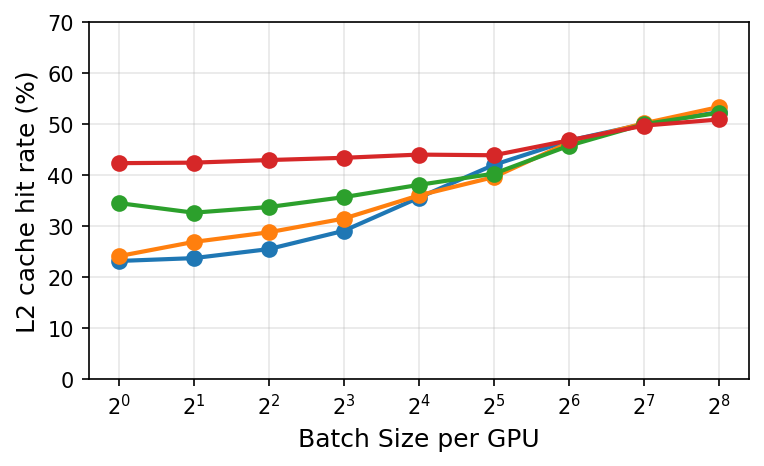} &
    \includegraphics[width=\imgsize]{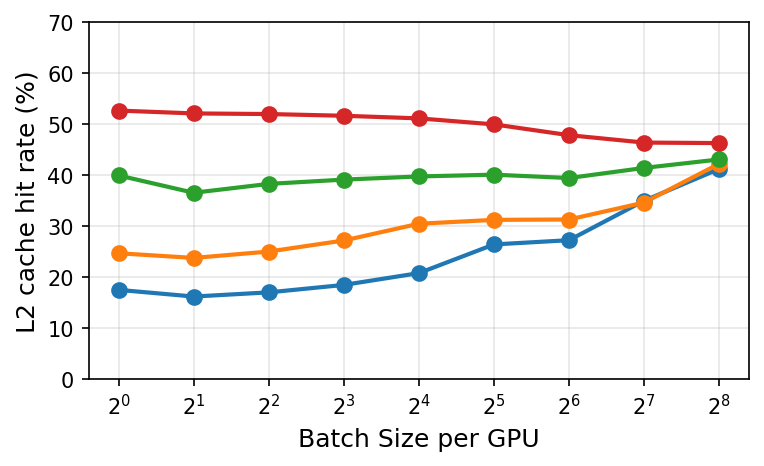} &
    \includegraphics[width=\imgsize]{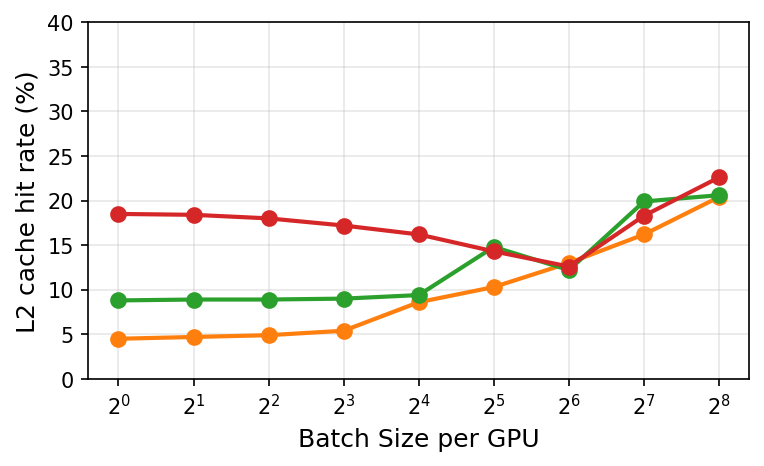} \\[-0.5ex]

    \includegraphics[width=\imgsize]{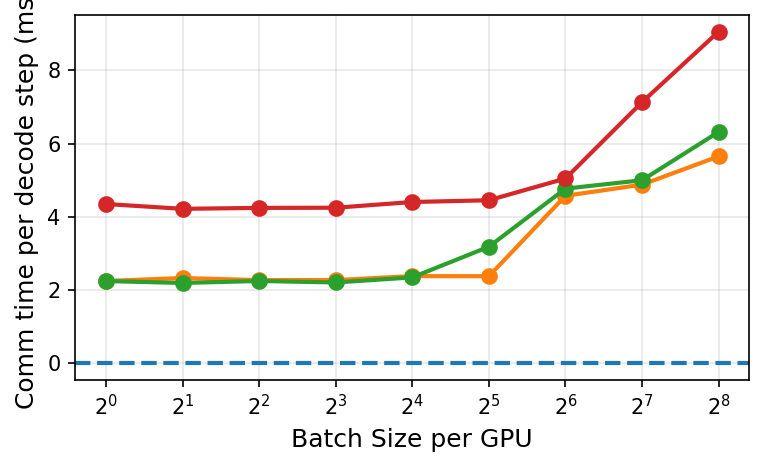} &
    \includegraphics[width=\imgsize]{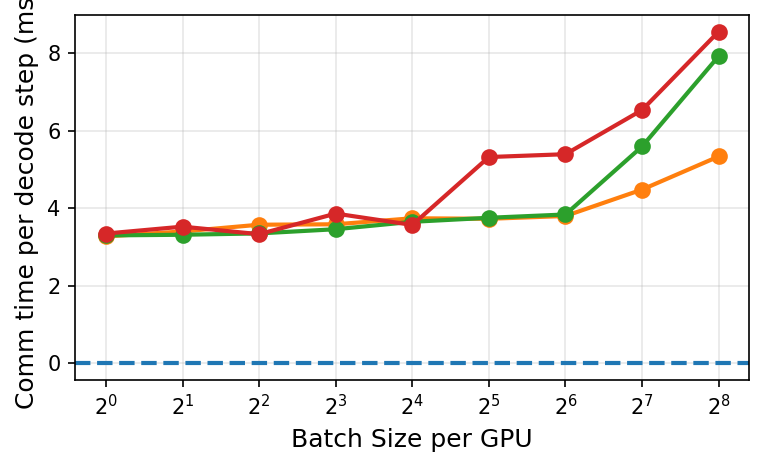} &
    \includegraphics[width=\imgsize]{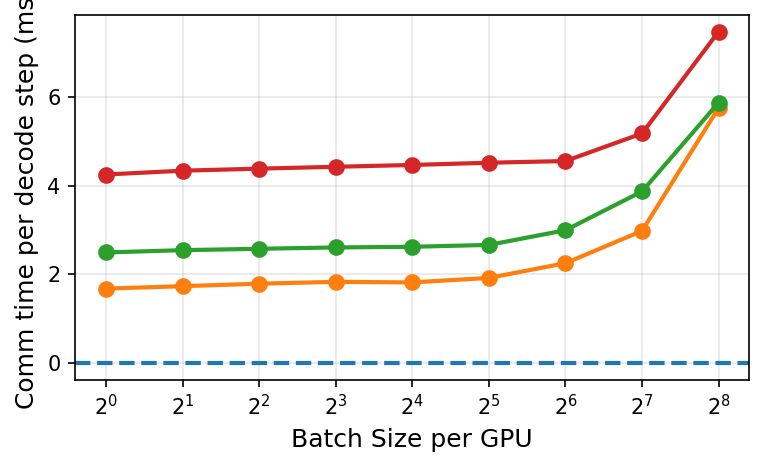}
    
  \end{tabular}
  \mycaption{fig:model_performance_grid}{Properties of Tensor parallelism}{Effect of Tensor Parallelism on 14B and 70B models across A100, H100, and B200 architectures via TTFT, Decode throughput, L2  Cache hit rate, and communication cost.}
\end{figure*}
\section{Background and Motivation}
\label{sec:motivation}

This section discusses the SLO-tier landscape in modern LLM serving, why tensor parallelism is an effective SLO-control surface, and why dynamic TP reconfiguration is necessary in practice.

\subsection{Low-Latency LLM and LLM SLO Tiers}
Large Language Models (LLMs) are employed across a spectrum of applications, each with distinct latency requirements. For instance, interactive copilots such as coding assistants require rapid responses to maintain user experience, often within milliseconds~\cite{sarathi}. Agentic workflows, including computer-use assistants (e.g., Claude Computer Use and OpenClaw), execute multi-step tasks with moderate but still user-visible latency expectations~\cite{anthropic-computer-use,openclaw-github}. Applications like content-creation tools and deep-research assistants can tolerate latencies ranging from seconds to minutes~\cite{mlc_serving}. At the relaxed end, recurring cron-style jobs and batch tasks (e.g., summarization and data analytics pipelines) can tolerate latencies up to hours~\cite{crontab-manpage,mlc_serving,openai2023batch}.

Production LLM systems already recognize these differing requirements, offering various tiers of subscription or API options tailored to user needs. For example, OpenAI, Anthropic Claude, and Google Gemini expose differentiated rate limits and subscription/pricing tiers~\cite{openai-rate-limits-guide,openai-api-pricing,anthropic-rate-limits-doc,anthropic-pricing,gemini-rate-limits-doc,gemini-subscriptions}. OpenAI also provides batch-processing modes for non-urgent queries at reduced cost~\cite{openai2023batch}. However, these offerings are primarily capacity and pricing controls rather than performance SLOs.

\begin{figure*}[htbp]
  \centering
  \begin{minipage}[t]{0.75\textwidth}
    \vspace{0pt}
    \centering
    \includegraphics[width=\textwidth]{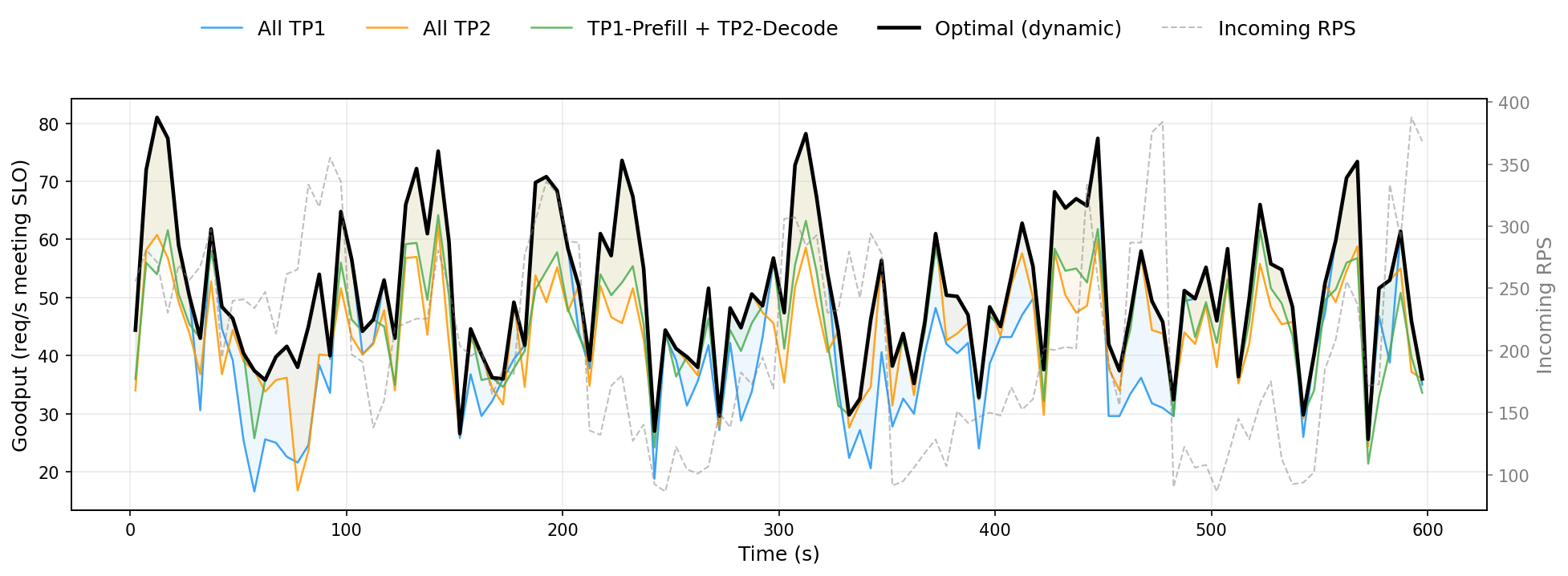}
  \end{minipage}
  \hfill
  \begin{minipage}[t]{0.24\textwidth}
    \vspace{0.9em}
    \centering
    \includegraphics[width=\textwidth]{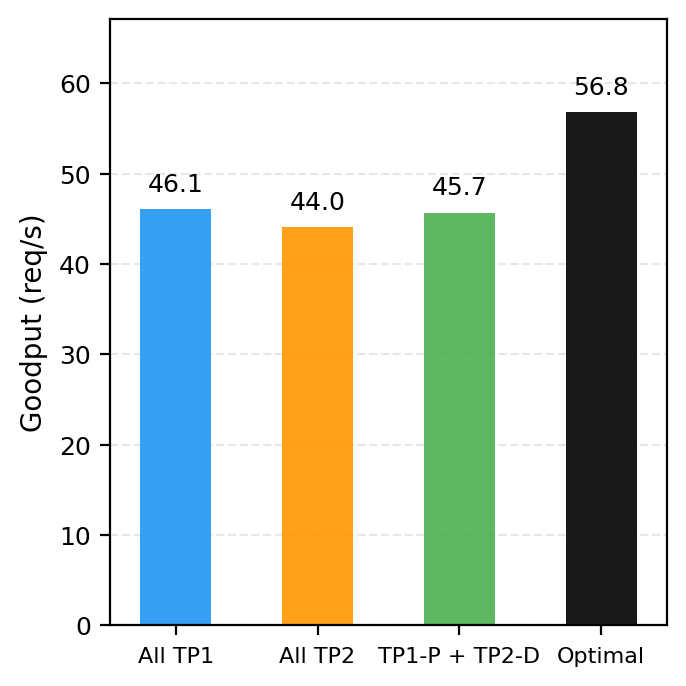}
  \end{minipage}
  \mycaption{fig:dynamic-goodput-row}{Static vs. Dynamic TP Goodput on the ServeGen Workload}{ Per-second goodput for three static TP
   baselines (All TP1, All TP2, TP1-Prefill + TP2-Decode) on A100, with
   incoming RPS overlaid (dashed, secondary axis). The black line (“Optimal”) is an oracle upper bound that selects the best configuration at each time step. No single static configuration dominates across time The bar chart reports aggregate goodput over
  the same 600s window.}
\end{figure*}

The concept of tiered SLOs itself has a longstanding presence in traditional data-center workloads~\cite{verma2015borg,burns2016borg-omega-kubernetes,silo2015,aws-cloudwatch-slo-doc,aws-cloudwatch-request-slo-2024}.
Different from traditional workloads, SLOs in LLM systems typically involve at least two distinct latency metrics: {\em Time to First Token (TTFT)} and {\em Time per Output Token (TPOT)}. TTFT measures the initial latency users experience before the first token is returned, impacting user-perceived responsiveness. TPOT assesses the latency to generate each output token, influencing the smoothness of streaming responses. 
%
Unlike traditional data-center SLOs, which often track a single end-to-end request latency, LLM serving must jointly satisfy two equally important objectives: fast response time (low TTFT) and sustained per-token generation speed (high TPOT), each with different resource bottlenecks and sensitivity to batch composition.
Thus, tiered SLOs in LLM serving pose new challenges in request scheduling, GPU resource management, and model forwarding mechanisms. 

\subsection{Why Tensor Parallelism for SLO}
\label{sec:motivate-tp}

Data parallelism (DP) and tensor parallelism (TP) are two common methods for scaling LLM inference across multiple GPUs. In DP, a model is replicated across GPUs. Each GPU receives a subset of the input batch and computes its outputs independent of the other workers. TP distributes model weights and the corresponding tensor computations across multiple GPUs. After each GPU's computation, its output must be synchronized via collective communication (all-reduce) with other workers. So far, TP has been primarily used for training or serving models whose weights are larger than what a single GPU can host.

We observe that TP can also be used to reduce prefill latency (TTFT) and increase decode throughput (1/TPOT). In Figure~\ref{fig:model_performance_grid}, we measure the TTFT and TPOT of different TP levels (1 to 8 for a DeepSeek-14B model running on 8 NVIDIA A100 GPUs and 2 to 8 for a llama3.1-70B model running on 8 NVIDIA H100/B200 GPUs) as prompt length (tokens) and batch size (number of requests) increase. We normalize the TPOT throughput by the number of GPUs (\ie, the TP level) to have a fair comparison. Our results show that higher TP level improves TTFT for all settings, especially as prompt lengths grows. This is expected because prefill applies large matrix operations across all prompt tokens; with higher TP, each GPU processes a smaller tensor shard, increasing effective compute and memory bandwidth per request and reducing TTFT when communication overhead remains smaller than the parallelism gain. 

TP's behavior in the decode phase is more subtle. When batch size is small, normalized per-GPU TPOT can be higher at higher TP levels, by up to 3 $\times$. This is counterintuitive because TP introduces cross-GPU communication, which is typically expected to reduce per-GPU efficiency. 

Digging into the underlying reasons, we find that smaller TP levels pay a higher cost to load larger weight matrices from global memory into per-SM memory, whereas higher TP levels load smaller matrix shards. We confirm this behavior by measuring the per-GPU L2 cache hit rate, as well as inter-GPU communication performance, in Figure~\ref{fig:model_performance_grid}. At small batch sizes, the intra-GPU memory-loading cost (\ie, L2 cache misses) dominates inter-GPU communication, yielding higher normalized per-GPU TPOT at higher TP. As batch size increases, the benefit shrinks: all TP settings approach the same intra-GPU memory-bandwidth ceiling, while higher TP levels incur higher communication overhead. 

This behavior persists across GPU types, even with newer generations like B200, and reflects fundamental properties of GPU architecture. Modern GPUs rely on a hierarchical memory system with limited on-chip cache capacity and high-bandwidth but high-latency off-chip memory, making many inference workloads, especially decode, sensitive to memory access patterns. At the same time, multi-GPU execution introduces communication overheads that depend on interconnect bandwidth. Although both memory and interconnect performance improve across generations, they scale under different constraints, and neither consistently dominates across all regimes. As a result, different TP levels will continue to exploit the tradeoff for future GPU generations.
Based on our TP study results, we propose to leverage TP to control SLOs based on batch size and inference stage.

\subsection{Why Dynamic TP for Tiered SLO}
\label{sec:dynamic-tp}

In tiered-SLO LLM serving, workload composition changes continuously over time: the fraction of requests in each SLO tier (\eg, strict interactive vs. relaxed background) can shift significantly across minutes and hours. Trace studies from Azure, Alibaba, and production-inspired LLM workloads report strong temporal variation in arrival rates, prompt/output characteristics, and service pressure across time \cite{dynamollm,burstgpt,azure-trace,alibaba-trace}. As this tier composition shifts, the TP setting that maximizes SLO-compliant goodput also shifts.

\begin{figure}[htbp]
  \centering
  \includegraphics[width=\linewidth]{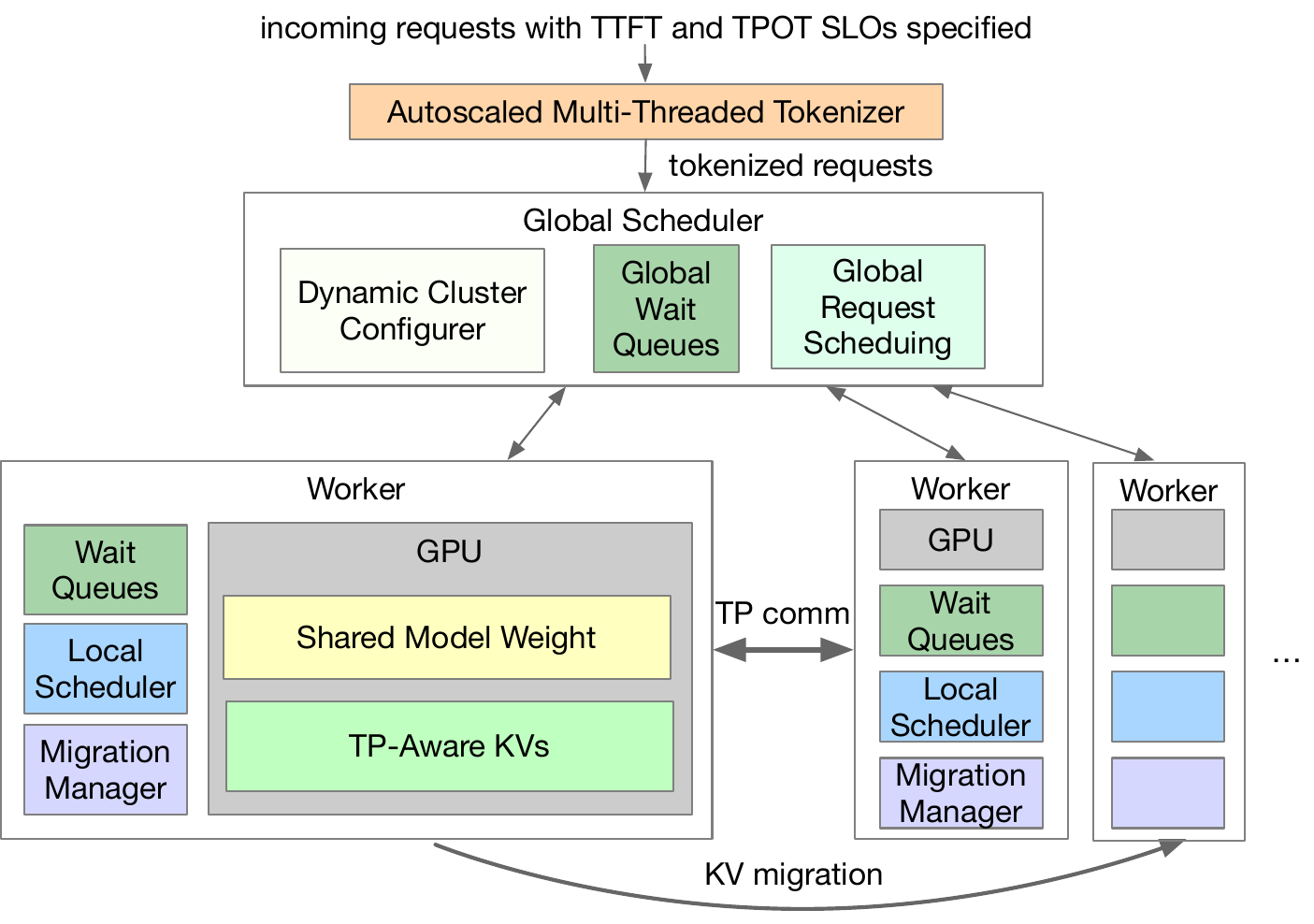}
  \caption{\textbf{\sys\ Architecture}}
  \label{fig-arch}
\end{figure}

Even within a single SLO tier, arrivals are bursty rather than smooth. User-facing traffic commonly exhibits micro-bursts from synchronized user behavior, workflow fan-out, and retry effects, producing short intervals where queueing delay dominates latency \cite{dynamollm,burstgpt}. A fixed cluster configuration cannot track these fast swings. 

We understand the need for dynamic TP configurations by studying the Alibaba LLM production trace~\cite{alibaba-trace}. Figure~\ref{fig:dynamic-goodput-row} shows the goodput when the A100 GPU cluster uses TP level 1, 2, and a mix of 1 and 2 (for prefill and decode, respectively), as well as an optimal TP setup that chooses the best TP configuration at runtime. The left figure is a timeline of a randomly sampled 10-minute window. As seen, each fixed configuration has distinct intervals of poor performance; and no single static configuration dominates over even a short 10-minute window. The right figure shows the overall goodput of the optimal dynamic TP setup and the fixed TP configurations, with the former being 23\% to 29\% higher.

  \section{\sys\ Design}
\label{sec:design}

This section first presents an overview of \sys's design.
We then introduce our mechanisms for efficient TP switching, discuss our GPU cluster configuration and request scheduling policy, and finally present our global and local scheduler designs.

\subsection{System Overview}
\label{sec:overview}


\sys\ is a distributed LLM serving system designed to maximize SLO-compliant throughput (goodput) under a fixed GPU budget in a multi-tier serving environment. Figure~\ref{fig-arch} shows its architecture. The design is centered on one idea: tensor parallelism (TP) should be treated as a runtime control surface for tiered-SLO serving, and to make it useful, TP switching should finish within milliseconds to track time-varying demand. \sys\ therefore combines low-overhead TP switching with goodput-aware cluster reconfiguration and SLO-aware request scheduling.


\boldunderpara{System model and control loop.}
\sys{} targets a homogeneous GPU cluster serving one model with multiple SLO tiers. Before serving begins, \sys{} profiles prefill/decode performance across TP levels, batch sizes, and sequence lengths. At runtime, the global scheduler combines these offline profiles with recent arrivals and TTFT/TPOT targets to pick a cluster configuration for each control window (default one second), then dispatches requests accordingly.

\boldunderpara{Design overview.}
\sys{} combines three components: (1) a goodput-aware global reconfiguration policy that selects TP and prefill/decode GPU allocation per tier; (2) low-overhead TP switching mechanisms, including reload-free TP weight switching and fast KV migration; and (3) SLO-aware request scheduling with global placement and per-GPU local batch control. Together, these components enable frequent adaptation while preserving fairness across tiers.

\subsection{Dynamic TP Mechanisms}
\label{sec:mechanism}

From our analysis in Section~\ref{sec:dynamic-tp}, maximizing goodput requires multiple TP levels and low-latency transition between them. 
Below, we describe our two efficient TP switching mechanisms that enable dynamic, frequent whole-GPU-pool reconfiguration (Section~\ref{sec:policy}).

{
\begin{figure}
\begin{center}
\centerline{\includegraphics[width=\columnwidth]{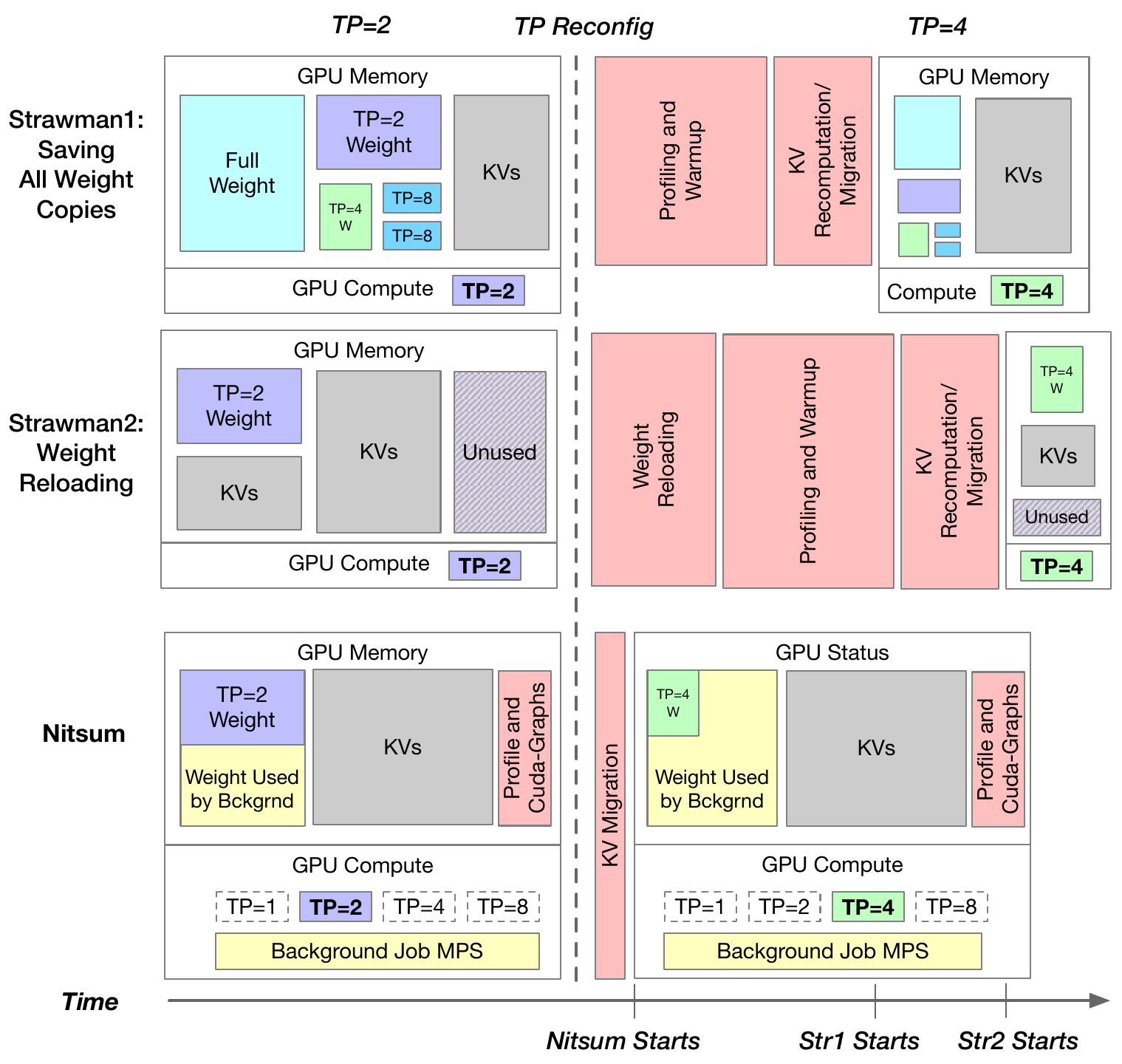}}
\mycaption{fig-tp}{Comparison of \sys\ and Straw-man TP Weight Loading.}
{
Straw-man 1 and 2 illustrate alternative solutions for dynamic TP for SLOs. Dashed boxes in the bottom row represent processes pre-allocated and kept warm by \sys; solid boxes represent running processes.
}
\end{center}
\end{figure}
}

\subsubsection{Zero-Overhead TP Weight Switching}
\label{sec:weight-layout}

With traditional TP, a model's weights are distributed across multiple GPUs, each occupying $1/N$ of each weight matrix for a TP level $N$. Thus, changing the TP level of a model means changing the weight matrix. A straightforward approach is to  as shown in Figure~\ref{fig-tp}. Doing so significantly slows TP reconfiguration, even when model weights are cached in CPU memory. For Qwen-14B, reconfiguration can take about 30 seconds, excluding profiling steps such as CUDA Graph capture and network initialization. A naive alternative is to store multiple versions of weights, one for each possible TP level. Having all possible weight versions in GPU memory avoids weight reloading, but storing them occupies memory that could otherwise be used to run more requests. For example, storing all TP-specific copies of a 13B Llama model requires 45.5 GB of GPU memory, which is about 56\% of an A100/H100's capacity.

{
\begin{figure}
\begin{center}
\centerline{\includegraphics[width=\columnwidth]{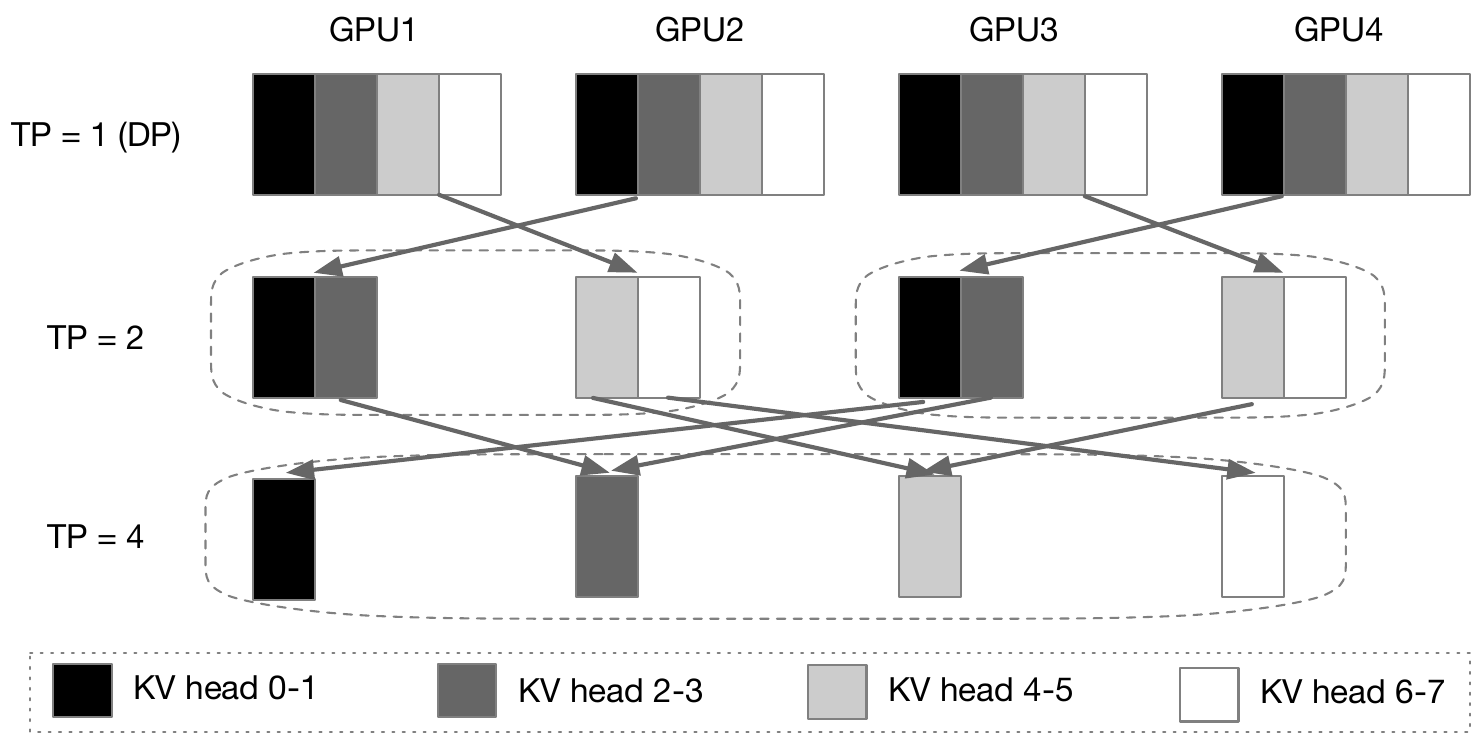}}
\mycaption{fig-tp-switch}{KV Conversion in \sys} 
{
When changing from TP 1 to TP 2 then to TP 4 on a cluster of four GPUs and 8 KV heads.
}
\end{center}
\end{figure}
}

Our solution is to keep one full copy of model weights on each GPU and select the TP-specific shard at execution time using customized kernels (Figure~\ref{fig-tp}). For example, under TP2, the first GPU in a TP group uses the first half of each weight matrix; under TP4, it uses the first quarter. This design eliminates weight reloads during TP reconfiguration, which would otherwise add substantial switching latency in dynamic workloads.

A natural concern is that some weights remain inactive at any given TP level. In practice, this overhead is often acceptable 
because under SLO-constrained serving, most server-grade GPUs are HBM-bandwidth-bound before they become memory-capacity-bound (although not to the extend that multiple copies of model weights can be stored). This is because KV footprints are often reduced by modern attention kernels~\cite{gqa,mla}, and SLOs further cap effective batch size. 
Overall, this weight-reuse design unifies weight storage and avoids weight reloading while remaining practical in real deployments.

A remaining challenge is that each TP level requires a different kernel for proper weight selection; naively initializing kernels at each TP switch introduces substantial setup and warmup overhead.
This is because existing systems typically pre-profile operation traces into fused launch units (cuda graphs in PyTorch) and apply compiler optimization (\texttt{torch.compile}) before serving starts to ensure efficient model forwarding performance. While these techniques improve steady-state forwarding efficiency, they make TP transitions slow (10seconds to 1min+ from our experiments), as illustrated by the strawman designs in Figure~\ref{fig-tp}.

Our solution is to decouple TP-switch latency from kernel preparation: we {\em pre-launch alive but inactive} execution processes for all candidate TP levels, each already specialized to its TP-specific weight layout. Concretely, we complete cuda-graph profiling and \texttt{torch.compile} optimization offline for every TP level before serving starts, then keep all corresponding GPU processes resident. At runtime, only the process for the active TP receives work, while the others remain hibernated with lightweight keep-alive signals. As a result, a TP switch activates an already-warm process, avoiding online profiling/compilation and enabling low-latency reconfiguration.

\subsubsection{Aggregated and Pipelined KV Migration}
\label{sec:migration}

When TP level changes, we need to ensure that both waiting requests and running requests' KVs are properly sent from a GPU to the appropriate destination GPUs.
Sending waiting requests is relatively straightforward and fast, as we only need to send the metadata and prompt of a waiting request from one GPU to multiple GPUs in the destination TP group or vice versa for TP level increase/decrease.

Handling running requests is more complex and critical to the overall performance. 
With different TP levels, KVs are evenly partitioned across GPUs on their head dimensions. 
As shown in Figure~\ref{fig-tp-switch}, assuming a total of eight heads, when switching from DP to TP 2 for the first two GPUs, we need to migrate KVs of heads 0 to 3 from GPU-2 to GPU-1 and KVs of heads 4 to 7 from GPU-1 to GPU-2 so that each GPU gets half of the heads for all the running requests on the two GPUs. Similarly, we gather KVs of heads 0 to 3 from GPU-4 to GPU-3 and heads 4 to 7 from GPU-3 to GPU-4. When switching from TP 2 to TP 4, we need to gather KV heads 0 and 1 from GPU-3 to GPU-1, KV heads 2 and 3 from GPU-1 and GPU-3 to GPU-2, and so on. 
The same process reverses when switching from high-level to lower-level TP.

To ensure that the correct KV heads for each running request are migrated to the right destinations, we first run a handshake across all participating GPUs to exchange the required migration metadata (e.g., request ID, KV head ID, and context length). After this metadata exchange, each GPU starts the actual KV transfer.

Existing KV migration techniques, used for prefill-decoding disaggregation, load balancing, defragmentation, etc., focus on hiding KV migration overhead by overlapping KV communication with foreground model forwarding~\cite{llumnix,zhong2024distserve}. Such live migration does not fit the need for TP switching, as the time it takes for the entire migration phase to finish is lengthy. Live migration requires batches of migrating newly generated KVs during the last batch of KV migration, similar to VM live migration~\cite{vm-live-migration-nsdi05}. Our evaluation shows that small batch of KV migration can take 300+ms under a typical setting for Llumnix~\cite{llumnix}. The lengthy live KV migration has two issues. First, it delays the time of starting the higher TP level which is needed to fulfill SLOs, impacting overall SLO attainment. Second, as \S\ref{sec:dynamic-tp} shows, micro-bursts are common and could last less than 10 seconds. By the time live KV migration finishes, workloads could already demand yet a new TP level.

{
\begin{figure}
\begin{center}
\centerline{\includegraphics[width=\columnwidth]{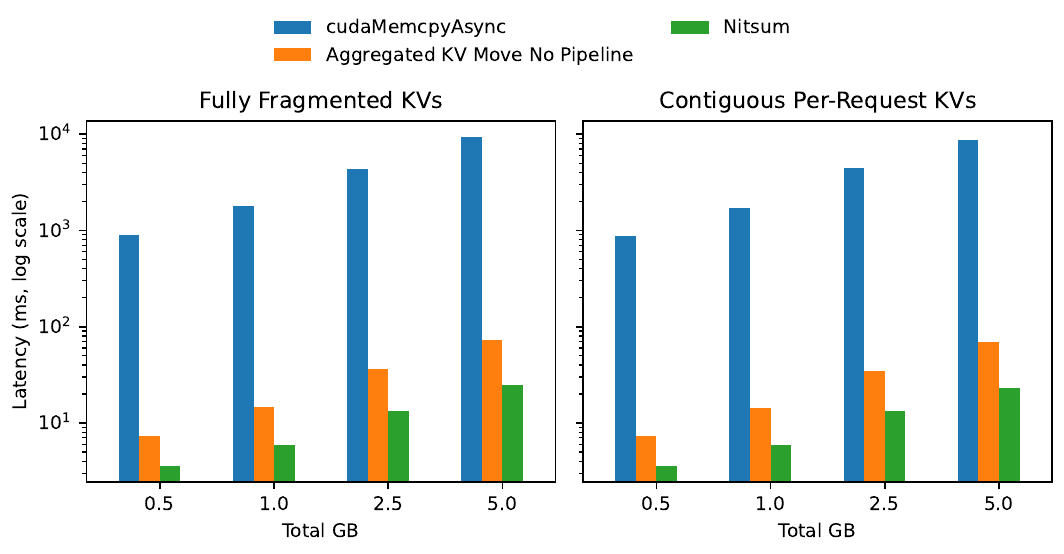}}
\mycaption{fig-kv-migration-microbench}{KV-Migration Latency Comparison.}
{
Transfer latency (log scale) across payload sizes for three strategies on fully fragmented and contiguous per-request KV layouts.}
\end{center}
\end{figure}
}

When digging into existing, vanilla KV migration (\eg, in Mooncake~\cite{mooncake}, Llumnix~\cite{llumnix}, DistServe~\cite{zhong2024distserve}), we find that the high KV migration latency comes from sending fragmented KV regions. 
Today's model inference systems~\cite{vllm-git-repo,sglang-git-repo} adopts PagedAttention~\cite{vLLM} and small page size to maximize GPU memory utilization and improve KV cache hit rate. Although benefitial for runtime model forwarding, small pages result in high KV fragmentation within a single request.
The KV context for a request can end up reside in a lot of discontigous memory space.
Standard memory-movement operations (\eg, cudaMemcpyAsync) issue a separate request for each memory page, resulting in many small transfers and high KV-migration latency.
Memory fragmentation is worth with higher TP levels, as KVs are shareded and each KV becomes smaller.

To mitigate these issues, our solution is to speed up KV migration process via a double-buffer aggregate and transfer mechanism.
Our customized KV migration kernel first copies fragmented KV regions to a contiguous buffer and then sends it directly to the other GPUs in the new TP formation. Instead of copying all KV regions into one giant contiguous buffer and then transmitting it, we use a pipelined mechanism to overlap memory copying and interconnect transmission. We use two relatively small temporary buffers. While we perform memory copy into the first buffer, the second buffer sends out the data copied in the previous stage. We then switch to sending the KV in the first buffer and copy memory to the second buffer. 

Figure~\ref{fig-kv-migration-microbench} plots our evaluation results for default cudaMemcpyAsync, an aggregated KV-move baseline, and \sys{}, which performs aggregation and transfer in a pipelined way. cudaMemcpyAsync takes about 0.88 to 9.25 seconds to migrate 0.5 to 5\GB\ of KV, or 4096 tokens to 40960 tokens for a fp16 LLama8B model. \sys{} reduces migration latency by 245$\times$ to 376$\times$ compared to cudaMemcpyAsync, with the resulting migration taking only 3.6 to 24.8 \ms.
Because of the minimal KV migration overhead, we adopt a stop-and-migrate approach where we pause model forwarding, migrate KVs, and then resume forwarding in the new TP level.

\subsection{Adaptive Configuration and Scheduling}
\label{sec:policy}

So far, we have presented how \sys{} achieves millisecond-level TP reconfiguration.
Now, we discuss the algorithm \sys{} uses to dynamically reconfigure a GPU cluster into different distributed serving plans (TP level, prefill/decode distribution), as well as how \sys{} schedules each individual request. Figure~\ref{fig-sched-algo} illustrates how \sys's configuration and scheduling system works at the high level.

\subsubsection{Distributed Serving Configuration}
\sys{}'s global scheduler performs fine-grained cluster reconfiguration at every time window (a configurable number default to one second).
Based on the request arrival rates of each SLO tiers in the past time window as well as offline profiled prefill/decode stats and TTFT/TPOT SLOs, the scheduler determines the cluster configuration for the next time window, including the number of GPUs used for prefill and decode stages in each SLO tier and their TP levels.

\begin{figure*}[htbp]
  \centering
  \includegraphics[width=0.85\textwidth]{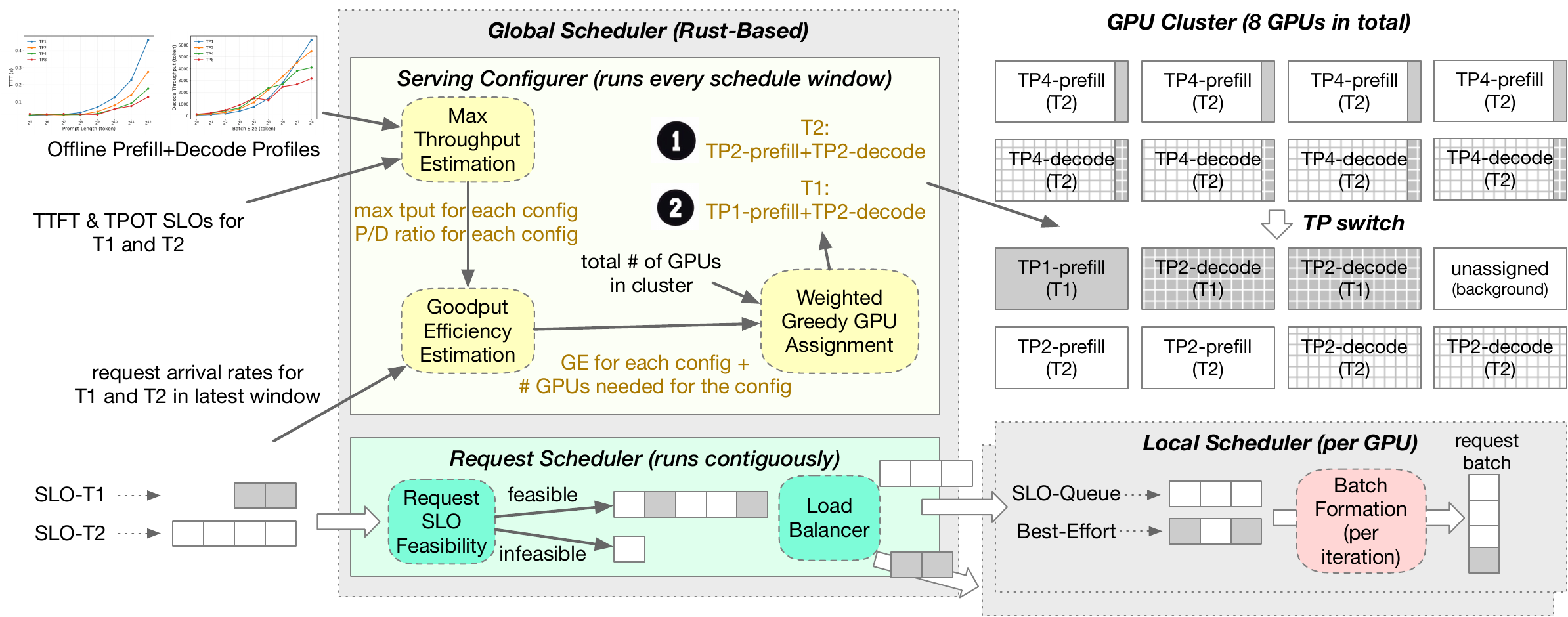}
  \mycaption{fig-sched-algo}{\sys\ Request Scheduling and Dynamic Serving Configuration.}
  {
  }
\end{figure*}

Our highlevel goal is to maximize the cluster's overall {\em goodput efficiency} (GE), defined as the SLO-compliant throughput normalized by the number of GPUs.
, while ensuring fairness across SLO tiers.
To achieve this goal, the global scheduler breaks down the task into two stages: (1) enumerate cluster configuration and estimate the goodput efficiency for each of them, (2) a weighted greedy algorithm to assign GPUs to the most efficient but still fair configuration.

\boldunderpara{Goodput efficiency estimation.}
To estimate goodput efficiency for a configuration for an SLO tier, we first estimate the maximum token throughput a configuration can achieve without violating TTFT and TPOT SLOs of the tier.

From Section~\ref{sec:motivate-tp}, we know each GPU type has a set of performance profiles for different TP levels and prefill/decode stages. These profiles do not change for different workloads or SLOs. Thus, we expect admins using \sys{} to perform offline profiling to acquire data points like those in Figure~\ref{fig:model_performance_grid} for each type of GPUs in their data center. 
For the TTFT/TPOT SLO of each tier ($tier\text{-}k$) and a chosen TP level ($TPi$), we can then deduct the maximum prefill and decode throughput, $THP_{tier\text{-}k}^{TPi}$ and $THD_{tier\text{-}k}^{TPi}$, by looking up the profiling results.

Since both TTFT and TPOT SLOs need to be met for a request to count towards goodput, we balance the prefill (P) and decode (D) resource ratio so that $P\times THP_{tier\text{-}k}^{TPi} = D \times THD_{tier\text{-}k}^{TPj}$, \ie, assigning $P \times TPi$ GPUs to prefill and $D \times TPj$ GPUs to decode.
After the prefill and decode stages are balanced, we can reduce the problem to only consider the max prefill throughput when calculating goodput efficiency going forward.

Next, we consider the incoming request rate ($rps_{tier\text{-}k}$) for each SLO tier and reduce the max prefill throughput to it if the rate is lower. The goodput effiency of a cluster configuration ($TPi,TPj$) and SLO $tier$-$k$ is 

\begin{equation}
    GE_{tier\text{-}k}^{TPi,TPj} = \frac{min(P \times THP_{tier\text{-}k}^{TPi},rps_{tier\text{-}k})}{P \times TPi + D \times TPj}
\end{equation}

\boldunderpara{Weighted greedy GPU resource assignment.}
With goodput efficiency calculated for each configuration, we then assign GPUs in the entire pool to configurations. However, the above problem is combinatorial: each tier can take multiple TP levels and GPU allocations, and the feasible configurations grow exponentially with the number of tiers and cluster scale, rendering it infeasible to solve exactly per control window.
We therefore adopt a greedy approximation that iteratively assigns GPUs to the configuration with the highest marginal gain. However, a naive greedy policy can starve tiers with lower efficiency.
To mitigate this issue, we introduce a weighted greedy assignment policy that prioritizes tiers with higher unmet demand. The weighted score, $WGE$, considers each tier's unmet incoming rps, i.e., $WGE_{tier\text{-}k}^{TPi,TPj} = GE_{tier\text{-}k}^{TPi,TPj} \times \frac{rps_{tier\text{-}k}}{served\text{-}rps_{tier\text{-}k}}$. We then conduct the greedy GPU assignment based on the $WGE$ scores.

\boldunderpara{Discussion.}
The success of the above dynamic cluster configuration algorithm hinges on two key factors. First, at each time window, we allow the entire pool to be reconfigured in any way that is the best for overall goodput efficiency. This is possible only because we achieve millisecond-level TP switching with our efficienct mechanisms (Section~\ref{sec:mechanism}). Second, we enumerate all possible configurations in the first step. We are able to finish this computation fast enough thanks to our multi-threaded Rust implementation of the global scheduler.

\subsubsection{SLO-Aware Request Scheduling}
The \sys\ global and local schedulers work together to perform SLO-aware request scheduling based on the current cluster configuration.
The global scheduler contiguously process incoming requests in the FCFS order.
If a request is a background one, the scheduler puts it in a background request queue and dispatches it to GPUs that accept background jobs (Section~\ref{sec:policy}) in a round-robin way.

For non-background requests, the global scheduler finds its SLO tier, the GPUs currently serving the tier, and the current remaining requests that these GPUs can still handle. For the last item, the global scheduler maintains a per-GPU SLO-compliant available serving bandwidth by deducting already assigned but not finished requests from the maximum throughput $THP_{tier\text{-}k}$. If adding the current request does not exceed this available serving bandwidth, the global scheduler labels it as a feasible request. Otherwise, the request is labeled infeasible (for achieving its SLO).

The global scheduler then dispatches the request to the most appropriate GPU. FOr a feasible request, it assigns it to the prefill GPU of its SLO tier that has the minimal current load (if multiple GPUs are serving as prefill workers for ther in the current configuration). For an infeasible request, we can assign it to any prefill GPU in the pool, with the hope that they will have residual resource now or in the future. Thus, the global scheduler spills infeasible requests to prefill GPUs in a round robin way.

\sys\ runs a local scheduler for each GPU. It maintains queues for feasible SLO requests, infeasible (best-effort) requests, and background requests. The local scheduler performs iteration-level scheduling to determine the batch formation for the next iteration. It limits the total batch size by the agreed upon $THP_{tier\text{-}k}$ or $THD_{tier\text{-}k}$ for its assigned $tire\text{-}k$ and then fills the batch with feasible requests. If the batch cannot be filled by then, the local scheduler fills the remaining slot with best-effort requests. When a request complete, the local scheduler informs the global scheduler so it can update its SLO-compliant available serving bandwidth for this GPU.

  \section{Evaluation Results}
We evaluate \textsc{\sys{}} to answer four key questions:
\begin{enumerate}
  \item Does \textsc{\sys{}} improve end-to-end SLO goodput?
  \item Can the benefit of \textsc{\sys{}} generalize to different workload/SLO settings?
  \item What components contribute to the gains?
  \item How well does \textsc{\sys{}} scale ?

\end{enumerate}

\label{sec:results}
We implemented \sys\ on top of SGLang~\cite{sglang} and rewrote its global scheduler, local scheduler, GPU kernels, and KV migration mechanism, with ~12K total source lines changed/added.

\begin{table}[t]
  \centering
  \small
  \begin{tabular}{lcccc}
  \toprule
   & \multicolumn{2}{c}{\textbf{Strict (Tier 1)}} & \multicolumn{2}{c}{\textbf{Loose (Tier 2)}} \\
  \cmidrule(lr){2-3}\cmidrule(lr){4-5}
  \textbf{Setup} & \textbf{TTFT} & \textbf{TPOT} & \textbf{TTFT} & \textbf{TPOT} \\
  \midrule
  Llama 8B A100 TP=1     & 500\,ms & 15\,ms & 500\,ms & 30\,ms \\
  Llama 8B H100 TP=1     & 300\,ms & 10\,ms & 300\,ms & 20\,ms \\
  Qwen 14B H100 TP=2              & 200\,ms & 10\,ms & 200\,ms & 15\,ms \\
  \bottomrule
  \end{tabular}
  \caption{Per-configuration SLOs on A100/H100 on Llama 8b/Qwen 14B}
  \label{tbl:slos}
  \end{table}

\boldunderpara{Baselines.}
We compare \sys{} against four baselines: (1) \textbf{Llumnix}~\cite{llumnix}: a state-of-the-art SLO-oriented distributed serving system based on vLLM that dynamically migrates requests across GPU instances to improve load balancing, reduce fragmentation, and prioritize high-SLO requests, (2) \textbf{Chiron}~\cite{chiron}: a hierarchical autoscaling system that uses local and global back pressure to adjust batch sizes and instance counts based on request SLOs, (3) \textbf{Split}: a set up that separates a GPU cluster into different groups with rejection, each for one SLO tier and running SGLang with a offline determined overall\footnote{the TP level that results in the highest goodput for the entire trace for each SLO tier} best TP level for that tier, (4) \textbf{SGLang (PD)}: a modified version of SGLang based on our prefill-decode disaggregation implementation, and (5) \textbf{SGLang}: the vanilla SGLang (default setting of no prefill-decode disaggregation).

\boldunderpara{Workloads.}
We evaluate our system using two sets of workloads.
The first set includes real traces from the \textbf{Azure} production LLM serving system~\cite{azure-trace}. These include model serving for coding tasks and for chats(roughly one hour of traffic each). The conversation trace includes requests from real customer chat with Azure's model backend, with an average prompt length of 1155 tokens, 211 output tokens, and an average arrival rate of 0.5 requests per second. The code trace represents copilot software engineering tasks served by Azure, with an average prompt length of 2048 tokens, 28 output tokens, and an average arrival rate of 2.3 requests per second. The original trace contains 1 hour wall-clock minutes of requests. To make experiment execution manageable, we select 10mins representative minutes for our evaluation. 

The second set is \textbf{ServeGen}~\cite{alibaba-trace}, a workload
generator calibrated to production LLM serving at Alibaba Cloud Model
Studio. We select two representative workloads from ServeGen's six
language workloads: conversation and code generation. The ServeGen
conversation trace has an average prompt length of $871$ tokens, $86$
output tokens, and an average arrival rate of $10.66$ requests per
second. The ServeGen code trace has an average prompt length of $912$
tokens, $148$ output tokens, and an average arrival rate of $11.94$
requests per second. Each trace spans a $10$-minute normalized
window.




\boldunderpara{Environments and models.}
We conduct experiments on GPU nodes in the RunPod GPU cloud~\cite{runpod}. Each node is equipped with 8 GPUs connected via NVLink. We evaluate two widely used GPUs: NVIDIA A100 and NVIDIA H100, both with 80 GB per GPU.
The A100 nodes provide 128 vCPUs and 2 TB of system memory, while the H100 nodes provide 224 vCPUs 2 TB of system memory. 

We use three representative open-source large language models: the Llama-3.1 8B model, Llama 3.1 70B, and the DeepSeek-R1-Distill-Qwen-14B model, all using 16-bit precision. These models differ in scale and architectural characteristics, allowing us to capture diverse compute and communication behaviors in LLM serving.
The 8B model fits in one GPU, while the 14B/70B model requires at least two GPUs.

\begin{figure*}[t]
    \centering
    \makebox[\textwidth][c]{%
      \includegraphics[width=1.2\textwidth]{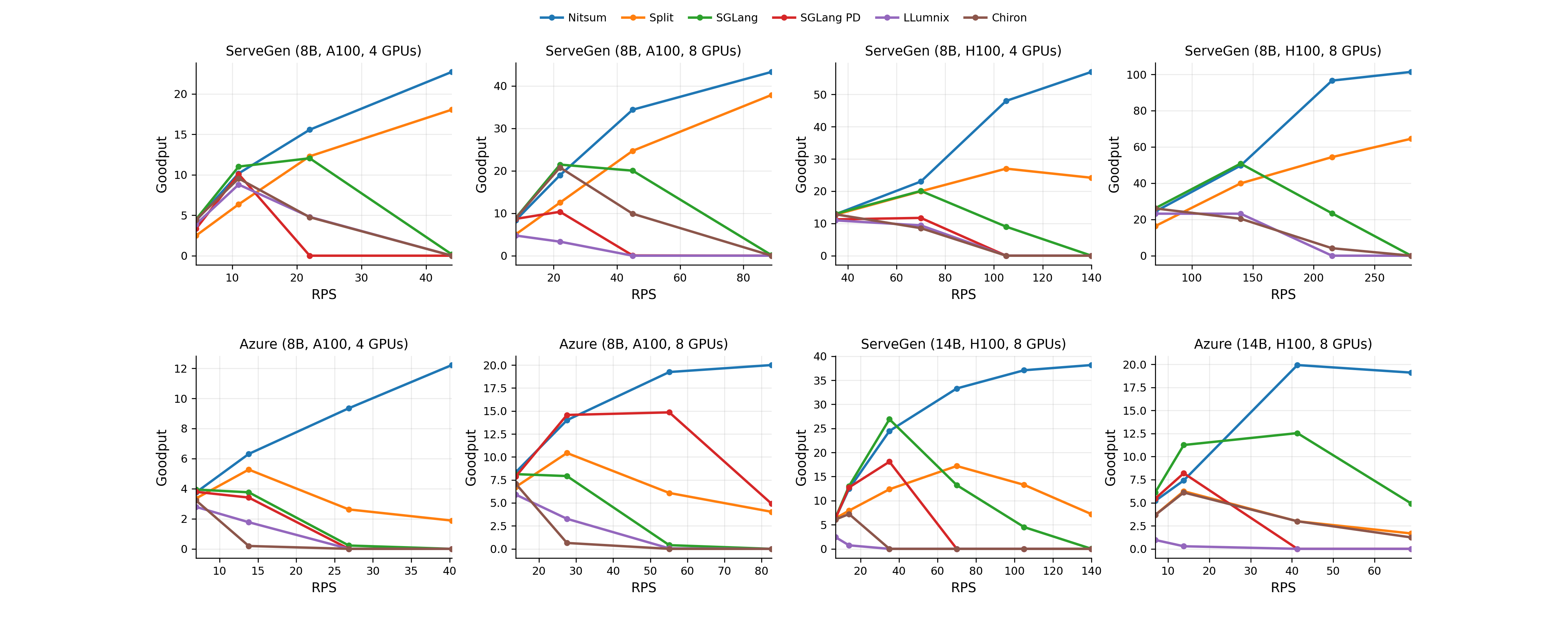}%
    }
    \vspace{0.1in}
    \mycaption{fig:e2e_results}{Goodput Results with Two Production Traces}{
      RPS shows incoming request per second. Goodput measured as requests meeting both TTFT and TPOT
  SLOs per second (higher is better). Results shown across two types of GPUs (A100, H100), two size of
   GPU cluster (4 and 8), two traces (ServeGen and Azure), and two model sizes (8B and 14B).
    }

    \par\vspace{10pt}

    \makebox[\textwidth][c]{
      \includegraphics[width=1.0\textwidth]{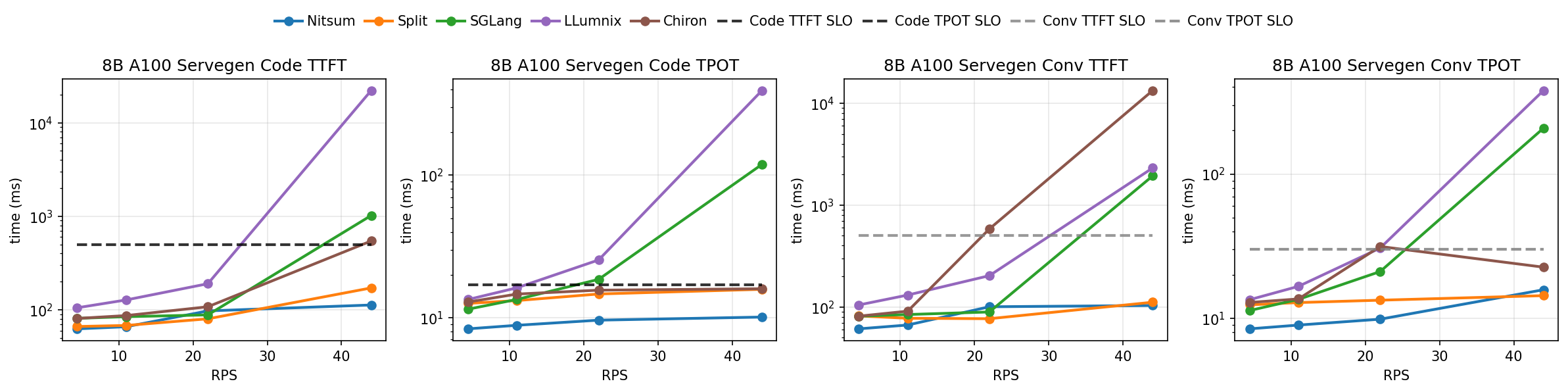}
    }
    \vspace{0.1in}
    \mycaption{fig:ttft-tpot-plot}{TTFT TPOT Raw Traces}{
      Median TTFT/TPOT collected from 8B A100 4 H100 across the code and conversation tiers on
  ServeGen workload.
    }
  \end{figure*}


\boldunderpara{SLO setups.}
 As no production SLOs are publicly known for model serving, we set TTFT and TPOT SLOs following the methodology used in SplitWise~\cite{patel2024splitwise}, by first measuring controlled microbenchmark performance and then scaling it by different factors for different tiers. Specifically, for each testing GPU cluster, we use the minimal TP level that a model fits and run one request at a time (\ie, batch size 1) using a workload's average token sizes. We record the measured average TTFT and TPOT time as the SLO for one tier (the strict, or high-priority, SLO). We then run the same setup but with high batch size (128) and measure the average TTFT and TPOT, as the SLO for a second tier (the relaxed, or low-priority, SLO).
Table~\ref{tbl:slos} summarizes the resulting SLOs.

\subsection{Overall Results}
We first present our end-to-end results of \sys{} and the five baselines running two models on two types of GPUs with two cluster size and the two workloads. 

\subsubsection{Workload Goodput}
Figure~\ref{fig:e2e_results} presents the goodput of eight settings as we vary the workload intensity. 
The X axis shows average requests per second injected to the system. Higher RPS means more intense load. 
The Y axis represent goodput, number of requests per second that meet both their TTFT and TPOT SLOs.

Overall, \sys{} achieves the highest goodput among all the systems, especially when the system load is high. \sys\ is also the only system that has consistently increasing goodput as system load increases. The gains stem from dynamically adapting TP levels and reallocating GPUs across tiers, particularly under highly variable workloads. 
As our adaptation responds to observed workload characteristics rather than fixed assumptions, \textsc{\sys{}} generalizes across both steady and bursty workloads, as well as different GPU types and cluster scales.

Most baseline systems collapse at some point for overall goodput. For example, all the baselines except for Split drops to 0 goodput beyond 40 incoming RPS for the 8B model running on 4 A100 GPUs (for both workloads). This is because they are either unable to adapt to the workload burstiness and are not properly routing/rejecting requests.

Among the baseline systems, Split performs better than Llumnix and Chiron, because it is able to isolate impact of the two SLOs. Default SGlang performs better in certain settings because it does not perform unnecessary reactions to the multi tier SLOs.


At low system load, all the systems perform similarly, suggesting that the amount of GPU is mostly more than enough for the load, regardless of the serving system.

\subsubsection{TTFT and TPOT Performance}
To understand where \sys{}'s overall goodput benefits come from, we measure the TTFT and TPOT time for all the systems. 
Figure~\ref{fig:ttft-tpot-plot} shows the median TTFT and TPOT for the ServeGen's Code and Convo workloads on the 8B model and 4 A100 GPUs. The SLOs used for these workloads are added as dashed flat lines, \ie, points below the lines mean meeting the respective SLOs. Across all the settings, \sys{} keeps both TTFT and TPOT below their SLOs, while most baselines violate one or both SLOs as RPS grows. 

Furthermore, \sys{}'s p90 TTFT and TPOT are still below the SLOs, while its p99 starts to violate SLOs for high RPS, as shown in Figure~\ref{fig:p90_ttft_tpot} and Figure \ref{fig:p99_ttft_tpot} in the Appendix. 


Notably, \sys{} achieves this goodput without sacrificing average latency: its average TTFT and TPOT are comparable to or better than the baselines, even for the low RPS regimes. 






\subsubsection*{TTFT and TPOT p90/p99 Performance}

\begin{figure*}[!t]
  \centering
  \begin{subfigure}{0.95\textwidth}
      \centering
      \includegraphics[width=\textwidth]{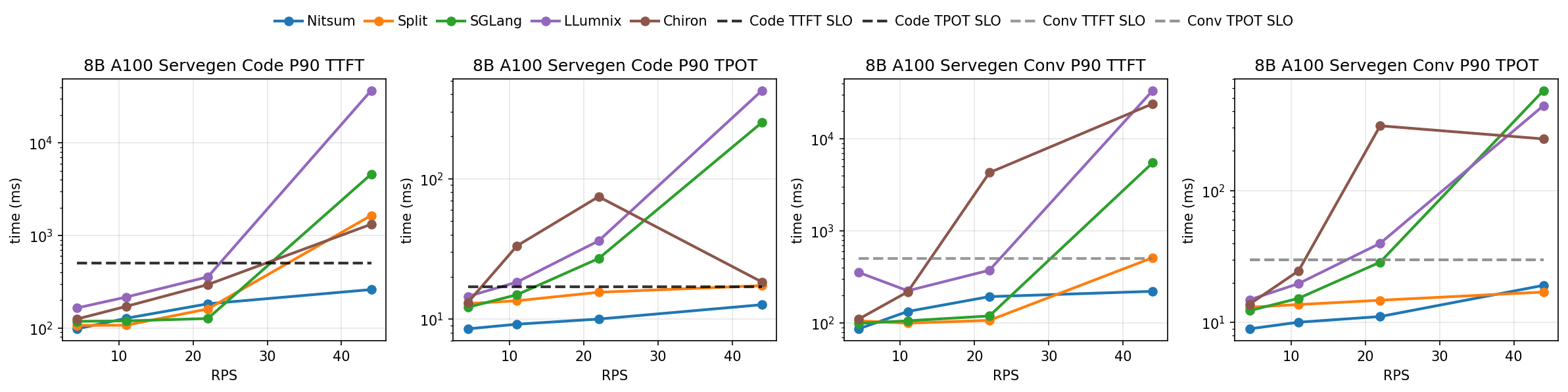}
      \caption{p90 TTFT and TPOT.}
      \label{fig:p90_ttft_tpot}
  \end{subfigure}

  \vspace{0.6em}

  \begin{subfigure}{0.95\textwidth}
      \centering
      \includegraphics[width=\textwidth]{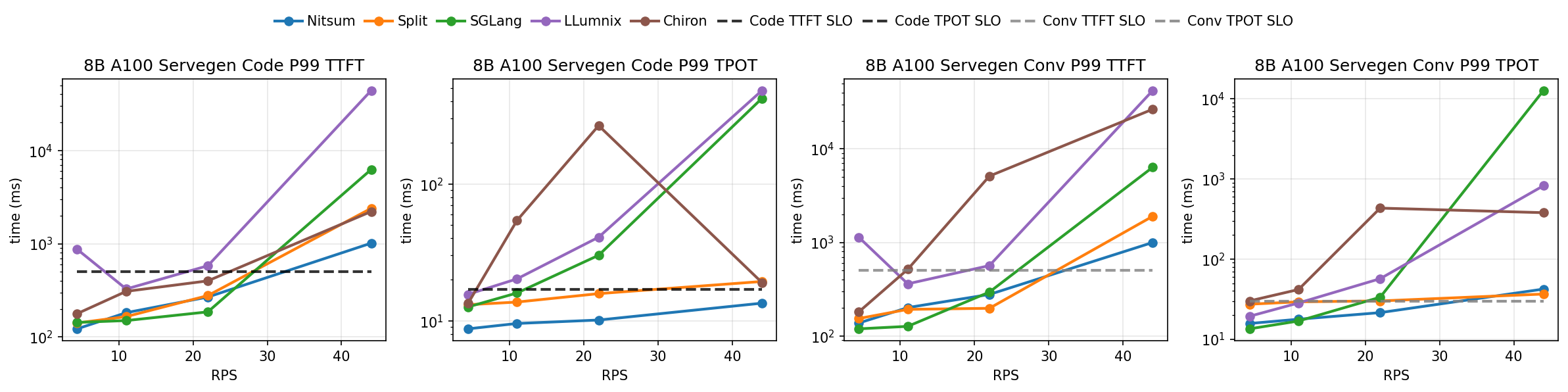}
      \caption{p99 TTFT and TPOT.}
      \label{fig:p99_ttft_tpot}
  \end{subfigure}

  \caption{\textbf{TTFT TPOT p90/p99 Raw Traces} \textit{Tail TTFT and TPOT under the ServeGen workload on 8B models with
4 A100 GPUs.} }
  \label{fig:appendix_tail_ttft_tpot}
\end{figure*}
We provide additional tail-latency results for TTFT and TPOT under the
ServeGen workload on 8B models with 4 A100 GPUs. We report both p90 in Figure \ref{fig:p90_ttft_tpot} and p99 in Figure \ref{fig:p99_ttft_tpot}.

\noindent
\textbf{Observations.}
Across both coding and conversation workloads, \sys{} consistently achieves
the lowest or comparable TTFT and TPOT at both p90 and p99. In contrast,
baseline systems degrade significantly as load increases. Systems with static
execution configurations experience growing queueing delays and reduced decode
efficiency, while systems relying on migration or autoscaling exhibit
unstable tail behavior under contention.

\noindent
\textbf{Tail amplification under load.}
The gap becomes more pronounced at p99, where even moderate inefficiencies in
scheduling or execution lead to large latency spikes. These results
demonstrate that \sys{} not only improves goodput, but also provides strong
tail-latency behavior by dynamically adapting TP level and prefill/decode
allocation.

\begin{figure*}[t]
    \centering
    \newcommand{\figheight}{3.2cm}
    \begin{minipage}[t]{0.44\textwidth}
      \centering
      \includegraphics[height=\figheight,width=\linewidth,keepaspectratio]{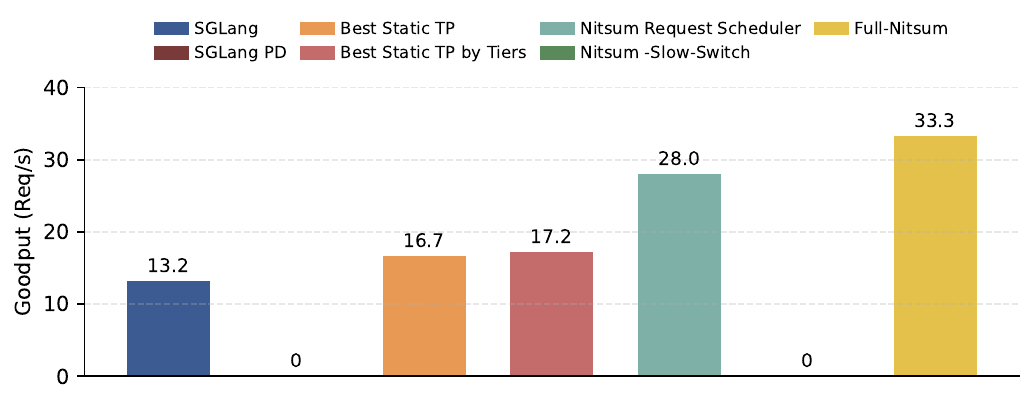}
      \mycaption{fig:ablation}{Ablation Study}{
        Progressively adding features from vanilla SGLang (leftmost) to full \sys. In certain case, adding a feature brings down goodput to zero.
      }
    \end{minipage}\hfill%
    \begin{minipage}[t]{0.26\textwidth}
      \centering

  \includegraphics[height=\figheight,width=\linewidth,keepaspectratio]{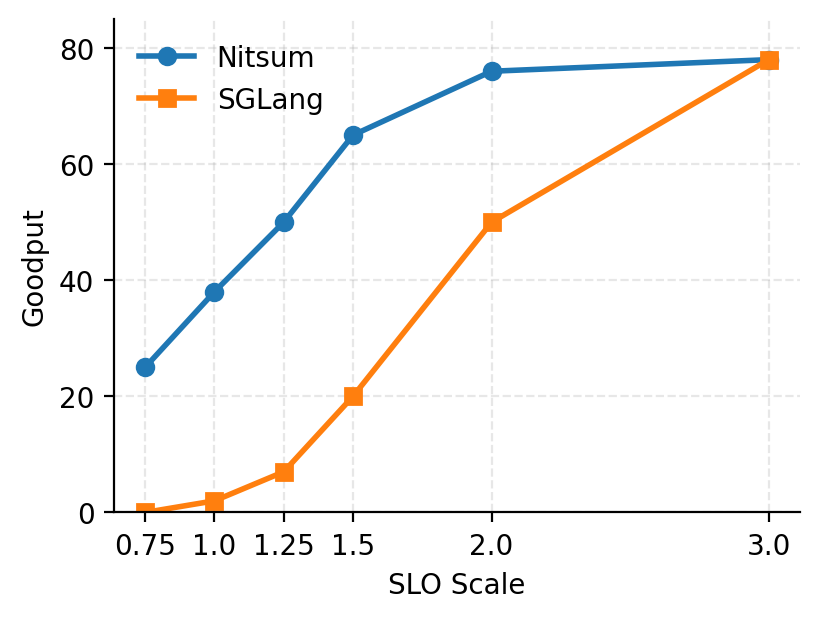}
      \mycaption{fig:slo-scaled}{Strict-Tier SLO vs.\ Goodput}{
        X axis shows a scale factor of SLOs in Table~\ref{tbl:slos}, smaller means tighter SLOs.
      }
    \end{minipage}\hfill%
    \begin{minipage}[t]{0.26\textwidth}
      \centering
      \includegraphics[height=\figheight,width=\linewidth,keepaspectratio]{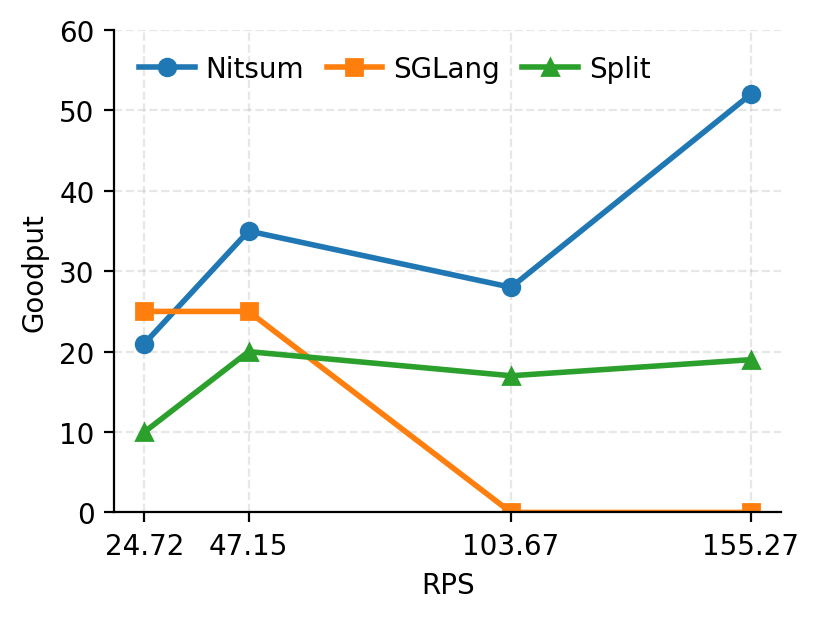}
      \mycaption{fig:multi-slo-relaxed}{Multi-Tier Relaxed SLOs}{
        Running three SLO tiers using the ServeGen workloads.
      }
    \end{minipage}
  \end{figure*}

\subsection{Deep Dive}

Now we provide a set of deep-dive experimental results that explain where \sys{}'s gain comes from and how \sys{} performs in different scenarios.

\subsubsection{Ablation Study}
To isolate the sources of \textsc{\sys{}}'s gains, we incrementally add its key mechanisms and measure workload goodput. Figure~\ref{fig:ablation} reports results for the 14B model on 8 H100 GPUs at 70 requests/s; other settings show similar trends.

Vanilla SGLang with a static TP-1 configuration and no prefill-decode disaggregation is SLO-agnostic and achieves only 13.2 req/s goodput. Enabling disaggregation in SGLang unexpectedly reduces goodput to zero. The static configuration mismatches prefill and decode capacity to the workload mix, causing one stage to overload and collapse.

Adding SLO awareness with a simple batch rule, which predicts whether each request can meet its SLO and defers those that cannot, and using the best static TP level for the trace raises goodput to 16.7 req/s. Further partitioning the cluster by SLO tier and using the best static TP per tier yields only a small gain, to 17.2 req/s. This shows that tier separation alone is insufficient.

Adding \textsc{\sys{}}'s SLO-aware scheduler, which combines goodput-aware batch composition, prefill/decode-aware assignment, rate limiting, and bin-packing placement, increases goodput to 28.0 req/s. This demonstrates the value of coordinated scheduling and resource-aware placement.

Dynamic TP requires TP reconfiguration and KV migration. When added with a naive weight layout and cudaMemcpyAsync-based KV migration, goodput again drops to zero because every TP switch incurs excessive overhead. This confirms that low-latency TP switching is essential for using dynamic TP to manage SLOs.

Finally, the full \sys{} system, with efficient dynamic TP reconfiguration, reaches 33.3 req/s goodput. Thus, dynamic parallelism adaptation provides the largest gains only when paired with efficient switching.

\subsubsection{Different Serving Targets}

To model different production systems, we perform a set of experiments to change SLO targets and add SLO tiers.

\boldunderpara{SLO strictness.}
So far, we have only experimented with one set of SLOs as in \autoref{tbl:slos}. To evaluate different SLO scenarios that production system may choose, we scale the TTFT and TPOT SLOs by a factor of 0.75 (more strict) to 3 (more relaxed) for the ServeGen workload and the 14B model using 8 H100 GPUs, as shown in ~\autoref{fig:slo-scaled}. \sys{} consistently outperforms SGLang across the entire range. The gains are largest at moderate SLO levels and diminish as SLOs become very loose (higher scale factor). This is because under moderately tight SLOs, different tiers favor different TP configurations, making dynamic adaptation effective. As SLOs relax further, the optimal choice converges to a uniform, low-TP configuration that maximizes throughput.

\boldunderpara{More SLO tiers.}
Our main experiments already include two SLO tiers. Here, we add a third workload from the ServeGen suite, with more relaxed SLO (600ms TTFT, 60ms TPOT), to evaluate behavior under increased tier diversity. As shown in ~\autoref{fig:multi-slo-relaxed}, Split is more stable than vanilla SGLang at high load, as resource isolation reduces cross-tier interference. However, at low load, Split underperforms SGLang because static partitioning prevents efficient resource multiplexing when some tiers have limited demand. In contrast, \sys{} consistently achieves the highest goodput across all regimes, as it enables space sharing across tiers and jointly selects the optimal resource allocation and TP levels based on the workload.



\subsubsection{System Scalability}

\boldunderpara{Scale to larger cluster}
We evaluate how goodput scales with the number of GPUs by varying the cluster size from 4 to 64 GPUs and measuring the maximum goodput in requests per second at which each system maintains at least 80\% overall SLO attainment. As shown in ~\autoref{fig:gpu-scaling}, \sys{} consistently achieves higher goodput across all scales and exhibits near-linear scaling as GPU count increases. To match \sys{}’s performance, baselines require 1.5–4x more GPUs.

\begin{figure*}[t]
  \centering
  \begin{minipage}[t]{0.32\textwidth}
    \centering
    \vspace{0pt}
    \includegraphics[width=\linewidth]{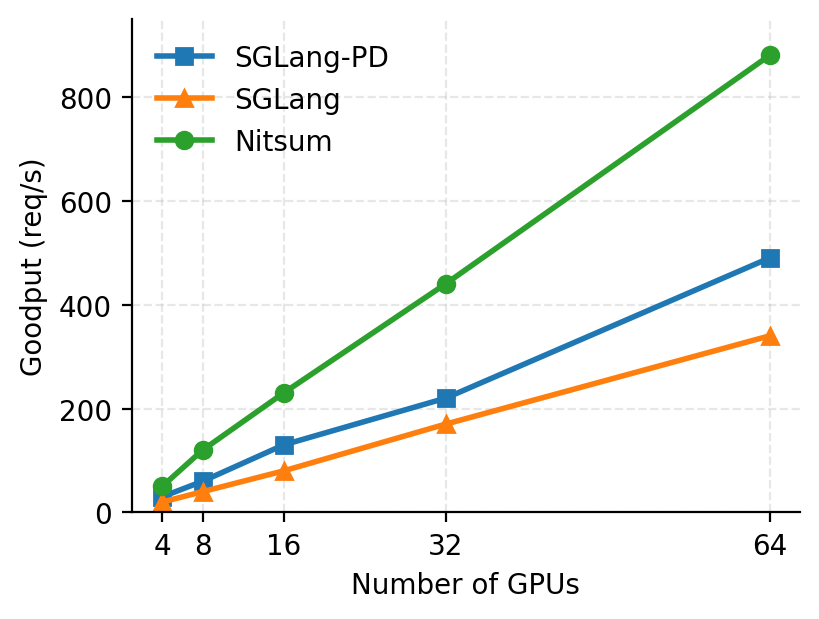}
    \mycaption{fig:gpu-scaling}{Goodput scalability with increasing GPU count}{}
  \end{minipage}
  \hfill
  \begin{minipage}[t]{0.32\textwidth}
    \centering
    \vspace{0pt}
    \includegraphics[width=\linewidth]{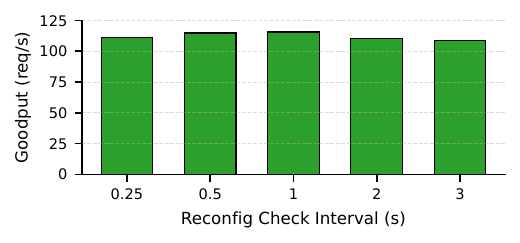}
    \mycaption{fig:reconfig_sensitivity}{Sensitivity to reconfiguration interval}{}
  \end{minipage}
  \hfill
  \begin{minipage}[t]{0.32\textwidth}
    \centering
    \vspace{0pt}
    \includegraphics[width=\linewidth]{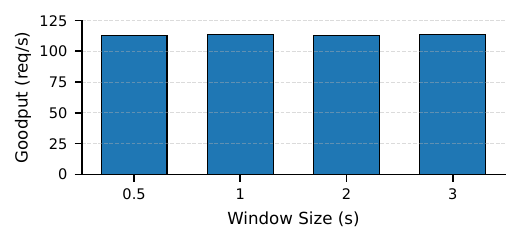}
    \mycaption{fig:window_sensitivity}{Sensitivity to monitoring window size}{}
  \end{minipage}
\end{figure*}

\boldunderpara{Scheduler scalability.}
\textsc{\sys{}}'s global scheduler is implemented as a lightweight single-process service in \textbf{Rust}, with an async HTTP frontend and a push-based dispatch pipeline to GPU workers. We evaluate the scheduler in a setting with 128 model replicas and send 50,000 tokenized requests in a batch to test the sustainable throughput of request routing. The batch is consumed in 3.1 seconds, corresponding to 16.1K requests per second throughput.

Importantly, control-plane operations are decoupled from the critical request path. In particular, tensor-parallel reconfiguration happens in a background thread and is computed by a greedy planner that searches only a controlled candidate space defined by a small fixed set of TP levels (e.g., TP1/2/4/8), rather than scaling with the full cluster size. As a result, the planning cost remains low even at a large scale: in our measurement with 128 GPUs and 4 request groups, reconfiguration planning takes only 2.49 ms on average. Overall, scheduler overhead scales with the available parallelism in the serving system without itself becoming a bottleneck in large deployments.


\subsubsection{Sensitivity Test}

We evaluate how \sys{}’s performance varies with key parameters to assess robustness and identify effective operating ranges.

\boldunderpara{Reconfiguration interval sensitivity.}
As shown in ~\autoref{fig:reconfig_sensitivity}, \sys{} achieves the best performance around 0.5–1 s, while both shorter and longer intervals slightly degrade. Very small intervals introduce unnecessary reconfiguration overhead, while large intervals react too slowly to workload changes. Overall, performance varies within ~6\%, indicating that \sys{} is robust to this knob, with a broad optimal region around sub-second intervals.

\boldunderpara{Profile-window sensitivity.}
In ~\autoref{fig:window_sensitivity}, we vary the time window used by the global scheduler to monitor per-request-group statistics (e.g., incoming RPS and prompt/decode-length distribution) and drive reconfiguration decisions. The result suggest this knob has little impact on the goodput.


  \section{Related Work}
\label{sec:related}

Recent LLM serving systems have increasingly focused on meeting heterogeneous service-level objectives (SLOs) under shared GPU resources. \sys{} builds on this line of work, but differs by treating model execution configuration as a runtime control surface.

\textsc{Llumnix}~\cite{llumnix} introduces a multi-instance scheduler that supports preemptive migration across GPU instances to balance load and reduce tail latency. However, Llumnix treats SLOs primarily through request placement and migration, without adapting the underlying model execution configuration. \sys{} complements Llumnix by using execution-level reconfiguration, specifically adaptive TP, to change request service times rather than only where requests execute.

\textsc{QLM}~\cite{onequeue} proposes a global scheduling queue with eviction, warm-starts, and model swapping to avoid head-of-line blocking. While effective in managing queue-level contention, it assumes static execution paths and does not exploit TP or MPS for deeper compute-level adaptation. In contrast, \sys{} dynamically selects TP degrees and co-locates slack-SLO requests via MPS to better utilize GPUs.

\textsc{SCOOT}~\cite{scoot} focuses on offline and online tuning of inference engine parameters (\eg, batch size and concurrency) to improve SLO attainment. It treats the inference stack as a black box and optimizes system-level hyperparameters. \sys{} differs by modifying the serving runtime to support dynamic parallelism adaptation and fast KV-cache migration.

\textsc{Chiron}~\cite{chiron} proposes hierarchical autoscaling for LLM serving, using SLO-aware backpressure to adjust batch sizes and allocate GPUs across interactive, mixed, and batch instances. While effective at improving SLO attainment and GPU efficiency, it primarily adapts request routing and serving capacity around fixed execution configurations. \sys{} differs by dynamically reconfiguring TP degrees, prefill/decode GPU allocation, and scheduling to adapt execution behavior under multi-tier SLO contention.

\section{Conclusion}

We presented \sys{}, a distributed LLM serving system that maximizes SLO-compliant throughput under a fixed GPU budget by dynamically adapting tensor parallelism (TP) and resource allocation.
Our key finding is that TP is not just a static deployment choice, but a runtime control surface that affects both TTFT and TPOT. To realize this, \sys{} enables low-overhead TP switching and goodput-aware reconfiguration, achieving consistent improvements over SoTA systems.

\sys{} is most beneficial under dynamic, multi-tier workloads where the optimal configuration changes over time; in more stable settings, a fixed TP configuration may suffice. It also incurs additional memory overhead and relies on workload estimates, which may limit effectiveness under highly constrained or unpredictable conditions.

Overall, this work highlights execution configuration as a dynamic resource in LLM serving systems, opening new directions for future heterogeneous-workload LLM serving.


  \bibliographystyle{plain}
  \bibliography{sysml,all-defs,references}

  \end{document}